# The Structure and Quantum Capacity of a Partially Degradable Quantum Channel

Laszlo Gyongyosi, *Member, IEEE*

[1] Quantum Technologies Laboratory, Department of Telecommunications
*Budapest University of Technology and Economics*
2 Magyar tudosok krt, Budapest, *H*-1117, Hungary

[2] Information Systems Research Group, Mathematics and Natural Sciences
*Hungarian Academy of Sciences*
Budapest, *H*-1518, Hungary

gyongyosi@hit.bme.hu

**Abstract**

The quantum capacity of degradable quantum channels has been proven to be additive. On the other hand, there is no general rule for the behavior of quantum capacity for anti-degradable quantum channels. We introduce the set of partially degradable (PD) quantum channels to answer the question of additivity of quantum capacity for a well-separable subset of anti-degradable channels. A quantum channel is partially degradable if the channel output can be used to simulate the degraded environment state. PD channels could exist both in the degradable, anti-degradable and conjugate degradable family. We define the term partial simulation, which is a clear benefit that arises from the structure of the complementary channel of a PD channel. We prove that the quantum capacity of an arbitrary dimensional PD channel is additive. We also demonstrate that better quantum data rates can be achieved over a PD channel in comparison to standard (non-PD) channels. Our results indicate that the partial degradability property can be exploited and yet still hold many benefits for quantum communications.

# 1  Introduction

The quantum capacity [5–7] of degradable quantum channels has been proven to be additive [8]. The fact that for a well-defined set of quantum channels the behavior of their quantum capacity is provably known makes the analysis easier in several practical cases. On the other hand, we have no fixed points from the field of anti-degradable channels. The analysis of these kinds of channels is much harder since the behavior of their quantum capacity depends on the properties of the actual map or the dimensions of the quantum channels [10]. Furthermore, from the analysis of the low-dimensional anti-degradable channels no relevant conclusions can be made on the high-dimensional anti-degradable channels. For a Hadamard channel [1–4], [11-12], the analysis of degradable channels can be made by studying the properties of its complementary channel (the channel between Alice and the environment). The complementary channel of a Hadamard channel is the entanglement-breaking channel and the classical and quantum capacity of this kind of structure is always additive [1-2], [10]. On the other hand, up to date we have no similar well defined structure in the world of anti-degradable channels.

The description of the quantum capacity of the anti-degradable channels would be much easier if we had found that a similar structure to the Hadamard channels exists for anti-degradable channels. We found that similar fixed points exist in the world of anti-degradable channels. This structure also equipped with a well defined complementary channel that determines the behavior of the quantum capacity of the channel and the structure holds this property in arbitrary dimensions. Like the Hadamard channels, which were playing the role of the base-stone in the degradable family thanks to their clear and well-defined structure and easy capacity evaluation process [9,13], our channel structure is aimed at serving the same role in the world of anti-degradable quantum channels. We called this family *partially degradable* (PD) channels. PD channels also could exists in the world of degradable channels, however this property trivially could not cause any change in the behavior of their quantum capacity, i.e., it will remain additive. On the other hand, the amount of information leaked to the environment is lower in comparison to those channels for which the partial degrading map does not exist. We also revealed that better quantum data rates can be achieved by PD channels in comparison to standard degradable or anti-degradable channels.

The PD channels have similar benefits in the world of anti-degradable channels like the Hadamard channels in the degradable set. The PD channels also have a clear and well-defined structure and offer several benefits for the evaluation of quantum capacity. The channel structure that makes the quantum capacity of a PD channel additive in arbitrary dimensions has to be equipped with an entanglement-binding [14,15] complementary channel. PD channels could be degradable (*degradable PD channels*) and anti-degradable (anti-degradable *PD channels*) ones, and also could be conjugate degradable (*conjugate-PD channels*) channels. We called these channels to the conjugate-PD channels, and formulate a subset in the PD channels. For a conjugate degradable channel, the environment state can be expressed from the channel output state up to a complex conjugation $\mathcal{C}$. The conjugate degradability property was introduced by Bradler *et al.* [1], and it was shown that there exist several finite-dimensional conjugate degradable channels. It was immediately recognized that these channels should have many useful benefits in quantum communications and paved the way for the introduction and definition of new channel sets that exist beyond the degradable and anti-degradable family, – *beyond the current boundaries*. The sets of *more capable* and *less noisy* quantum channels were introduced by Watanabe [2], and it was further demonstrated that other, well-characterized channel sets could be defined beyond the already existing sets. Here, we continue the way and make a step further to extend the possibilities for constructing a more general set motivated by the results obtained in [1].

This paper is organized as follows. In Section 2, the preliminaries are summarized. Section 3 discusses the theorems and proofs, while Section 4 reveals some examples. Finally, Section 5 concludes the paper. A code construction scheme and the performance analysis of PD channels are included in the Supplemental Information.

## 2 Preliminaries

### 2.1 Basic Definitions and Notations

In this section we summarize the basic definitions and notations.

**Quantum channel**

A quantum channel $\mathcal{N}$ is a CPTP (Completely Positive Trace Preserving) map. The action of a $d$ dimensional noisy quantum channel $\mathcal{N}$ on input density matrix $\rho$ can be expressed in the Kraus representation as follows:

$$\mathcal{N}(\rho) = \sigma = \sum_i N_i \rho N_i^\dagger, \tag{1}$$

where $\sigma$ is the channel output, while $N_i$ are the Kraus operators with $\sum_i N_i^\dagger N_i = I$.

**Isometric extension**

The isometric extension of a noisy quantum channel $\mathcal{N}$ with input $A$, output $B$ and environment $E$ is denoted by $U_{\mathcal{N}}^{A' \to BE}$ and expressed as

$$U_{\mathcal{N}}^{A' \to BE} = \sum_i N_i^{A \to B} \otimes |i\rangle_E, \qquad (2)$$

where $N_i^{A \to B}$ are the Kraus operators and the set $\{|i\rangle_E\}$ is an orthonormal set of states. Defining isometry $U_{\mathcal{N}}$ of noisy channel $\mathcal{N}$, for an input density matrix $\rho$ the conjugation of the input system is expressed as

$$U_{\mathcal{N}} \rho U_{\mathcal{N}}^{\dagger} = \sum_{i,j} \left( N_i \rho N_j^{\dagger} \right)_B \otimes |i\rangle\langle j|_E, \qquad (3)$$

where $U_{\mathcal{N}}^{\dagger} U_{\mathcal{N}} = I$.

**Logical channels**

The isometric extension of a noisy channel $\mathcal{N}$ makes possible to view the channel as a logical channel between Alice and Bob, and a logical channel between Alice and the environment called the *complementary channel*. The logical channels of $\mathcal{N}$ are denoted by $\mathcal{N}_{AB}$ and $\mathcal{N}_{AE}$, and expressed by isometry $U_{\mathcal{N}}$ as follows:

$$\mathcal{N}_{AB}(\rho) = Tr_E \left\{ U_{\mathcal{N}} \rho U_{\mathcal{N}}^{\dagger} \right\}, \qquad (4)$$

and

$$\mathcal{N}_{AE}(\rho) = Tr_B \left\{ U_{\mathcal{N}} \rho U_{\mathcal{N}}^{\dagger} \right\}, \qquad (5)$$

where $Tr_i(\cdot)$ is the partial trace operation, and $U_{\mathcal{N}} \rho U_{\mathcal{N}}^{\dagger} = \sum_{i,j} \left( N_i \rho N_j^{\dagger} \right)_B \otimes |i\rangle\langle j|_E$ with $U_{\mathcal{N}}^{\dagger} U_{\mathcal{N}} = I$.

Using Kraus operators $N_i$, the complementary channel $\mathcal{N}_{AE}$ can be rewritten as

$$\mathcal{N}_{AE}(\rho) = \sum_{i,j} \left\{ N_i \rho N_j^{\dagger} \right\} |i\rangle\langle j|_E. \qquad (6)$$

**Output and environment dimensions**

Let $\rho$ be the input system with dimension

$$d_A = \dim\{\rho\}. \qquad (7)$$

The dimension of channel output system $B$ of $\mathcal{N}$ using logical channel $\mathcal{N}_{AB}$ is denoted by

$$d_B = \dim\{\sigma_B\} = \dim\{\mathcal{N}_{AB}(\rho)\}. \tag{8}$$

The dimension of the environment state $E$ is expressed by the output of the complementary channel $\mathcal{N}_{AE}$ as

$$d_E = \dim\{\sigma_E\} = \dim\{\mathcal{N}_{AE}(\rho)\}. \tag{9}$$

For channel $\mathcal{N}$ the Choi rank can be expressed as the rank of the channel's Choi Jamiolkoswski state (CJ-state) $\sigma = (I \otimes \mathcal{N})\rho$, where the input system is characterized as follows

$$\rho = |\psi\rangle\langle\psi|, \ |\psi\rangle = \tfrac{1}{\sqrt{d_A}}\sum_{i=0}^{d_A-1}|i\rangle|i\rangle, \tag{10}$$

where $d_A$ is the dimension of input system $\rho$. For the rank of the CJ state the following relation holds:

$$rank\{\sigma = (I \otimes \mathcal{N})\rho\} = \min\dim\{\sigma_E\} = \min\dim\{\mathcal{N}_{AE}(\rho)\} = \min(d_E), \tag{11}$$

where $d_E$ is the dimension of the environment state $\sigma_E$ outputted by the complementary channel $\mathcal{N}_{AE}$.

For any $\mathcal{N}$ with logical channels $\mathcal{N}_{AB}$ and $\mathcal{N}_{AE}$, between the dimensions of input, output and environment states the following relation holds:

$$d_A d_B \geq d_E. \tag{12}$$

The CPTP maps $\mathcal{N}, \mathcal{N}_{AB}$ and $\mathcal{N}_{AE}$ can be rewritten as a map between $d_i$ dimensional matrices $M_{d_i}$ as follows:

$$\mathcal{N} : M_{d_A} \mapsto M_{d_B}, \tag{13}$$

and

$$\begin{aligned}\mathcal{N}_{AB} &: M_{d_A} \mapsto M_{d_B}, \\ \mathcal{N}_{AE} &: M_{d_A} \mapsto M_{d_E}.\end{aligned} \tag{14}$$

**Degrading map**

A degrading map $\mathcal{D}$ is a CPTP map that can be used to simulate the map of logical channel $\mathcal{N}_{AB}$ or the map of the complementary channel $\mathcal{N}_{AE}$. If the environment $\sigma_E = \mathcal{N}_{AE}(\rho)$ of the complementary channel $\mathcal{N}_{AE}$ can be simulated by a degrading map $\mathcal{D}^{B\to E}$ from the output $\sigma_B = \mathcal{N}_{AB}(\rho)$, then channel $\mathcal{N}$ is *degradable* and for all input systems $\forall \rho$:

$$\mathcal{N}_{AB}(\rho) \circ \mathcal{D}^{B\to E} = \mathcal{N}_{AE}(\rho). \tag{15}$$

*(Note:* The channel composition will be denoted in the order of $\mathcal{N} \circ \mathcal{D}$ throughout.) In other words, for a degradable channel $\mathcal{N}$, the channel output $B$ can be used to simulate the environment state $E$, because the noise of channel $\mathcal{N}_{AB}$ is lower than the noise of $\mathcal{N}_{AE}$.

If channel $\mathcal{N}$ is *anti-degradable* then the output system $\sigma_B = \mathcal{N}_{AB}(\rho)$ can be simulated by environment state $\sigma_E = \mathcal{N}_{AE}(\rho)$ for all input systems $\forall \rho$ as follows:

$$\mathcal{N}_{AE}(\rho) \circ \mathcal{D}^{E \to B} = \mathcal{N}_{AB}(\rho), \tag{16}$$

where $\mathcal{D}^{E \to B}$ is the degrading map, which degrades the environment state to the channel output state.

**Conditions of degradability**

Assuming logical channels $\mathcal{N}_{AB}$ and $\mathcal{N}_{AE}$ of $\mathcal{N}$, a necessary condition for degradability of $\mathcal{N}$ is as follows:

$$\ker\{\sigma_B = \mathcal{N}_{AB}(\rho)\} \subseteq \ker\{\sigma_B = \mathcal{N}_{AE}(\rho)\}, \tag{17}$$

where $\ker\{\cdot\}$ stands for the kernel of channel output density matrix. It follows that for $d_B \leq d_A$, the degradability requires also another condition on degrading map $\mathcal{D}^{B \to E}$, namely:

$$\mathcal{D}^{B \to E} = \mathcal{N}_{AB}^{-1} \circ \mathcal{N}_{AE} \tag{18}$$

on $\ker\{\sigma_B = \mathcal{N}_{AB}(\rho)\}^\perp$ has to be satisfied, where $\mathcal{N}_{AB}^{-1}$ is the inverse of channel $\mathcal{N}_{AB}$. If $d_B > d_A$, then $\mathcal{N}_{AB}^{-1}$ has no right inverse and $\mathcal{D}^{B \to E}$ cannot be unique [10].

For an anti-degradable channel the condition on $\mathcal{D}^{E \to B}$ is as follows:

$$\mathcal{D}^{E \to B} = \mathcal{N}_{AE}^{-1} \circ \mathcal{N}_{AB}. \tag{19}$$

**Entanglement-binding channels**

Channel $\mathcal{N}$ is called entanglement-binding if it has $Q(\mathcal{N}) = Q_2(\mathcal{N}) = 0$, where $Q_2(\mathcal{N})$ stands for the LOCC-assisted quantum capacity, and outputs a bound-entangled system $\sigma$. Assuming an entanglement-binding channel $\mathcal{N}: M_{d_A} \mapsto M_{d_B}$ with input system $|\Psi_{AA'}\rangle$ that is half $A'$ of the $d_{AA'}$ denoted by $d_A$ dimensional maximally entangled system

$$\rho_{AA'} = |\Psi_{AA'}\rangle\langle\Psi_{AA'}| : M_{d_A} \otimes M_{d_A}, \tag{20}$$

where

$$\left|\Psi_{AA'}\right\rangle = \tfrac{1}{\sqrt{d_A}}\sum_{i=1}^{d_A}\left|i\right\rangle\left|i\right\rangle, \tag{21}$$

the output of the channel is a $d_B$ dimensional bound-entangled system $\sigma$ as follows:

$$\sigma = (I \otimes \mathcal{N})\left|\Psi_{AA'}\right\rangle \tag{22}$$

which state is acting over the space of $\mathbb{C}^{d_A \cdot d_B}$.

For the input and output dimensions of any entanglement-binding channel $\mathcal{N}$ the following condition has to hold:

$$d_A d_B > 6 \tag{23}$$

otherwise bound entanglement could not exist. In other words for any entanglement-binding channel follows that is has to be a qudit channel, i.e.,

$$d_A = \dim\{\rho\} > 2, \tag{24}$$

and

$$d_B = \dim\{\sigma_B\} = \dim\{\mathcal{N}_{AB}(\rho)\} > 2. \tag{25}$$

**Entanglement-binding complementary channel**

In relation of an entanglement-binding complementary channel $\mathcal{N}_{AE}$ with input dimension $d_A$ and output dimension $d_E$ these condition as follows:

$$d_A = \dim\{\rho\} > 2, \tag{26}$$

and

$$d_E = \dim\{\sigma_E\} = \dim\{\mathcal{N}_{AE}(\rho)\} > 2, \tag{27}$$

i.e., $\mathcal{N}_{AE}$ also must be a qudit channel to satisfy the condition $d_A d_E > 6$.

**Bound-entangled state**

Any bound-entangled state can be expressed in the following form

$$\sigma = p\left|\Psi_{AA'}\right\rangle\left\langle\Psi_{AA'}\right|_r + (1-p)\tfrac{1}{r^2-1}(I - \left|\Psi_{AA'}\right\rangle\left\langle\Psi_{AA'}\right|), \tag{28}$$

where $\left|\Psi_{AA'}\right\rangle\left\langle\Psi_{AA'}\right|_r$ is a maximally entangled state with Schmidt rank $r$, and $p \leq \tfrac{1}{r}$.

Let $\left|\Psi_{AA'}\right\rangle$ be a $d_A = 3$ dimensional maximally entangled input system

$$\left|\Psi_{AA'}\right\rangle = \tfrac{1}{\sqrt{3}}\sum_{i=0}^{2}\left|i\right\rangle\left|i\right\rangle, \tag{29}$$

and let $\mathcal{N}$ be an entanglement-binding channel defined by the CPTP map [14] as follows:

$$\mathcal{N}(\rho) = \tfrac{2}{7}(\rho) + \tfrac{\alpha}{7}\sum_{k=1}^{3}|j\rangle\langle k|(\rho)|k\rangle\langle j| + \tfrac{5-\alpha}{7}\sum_{k=1}^{3}|l\rangle\langle k|(\rho)|k\rangle\langle l|, \qquad (30)$$

where $3 < \alpha \leq 4$, and

$$j = (k+1) \bmod 3, \qquad (31)$$

$$l = (k-1) \bmod 3. \qquad (32)$$

The output of this entanglement-binding channel $\mathcal{N}$ is the system $\sigma = (I \otimes \mathcal{N})|\Psi_{AA'}\rangle$ over $\mathbb{C}^{d_A \cdot d_B}$ expressed by

$$\sigma_\alpha = \tfrac{2}{7}|\Psi_{AA'}\rangle + \tfrac{\alpha}{7}\sigma_+ + \tfrac{5-\alpha}{7}\sigma_-, \qquad (33)$$

where $3 < \alpha \leq 4$ and

$$\sigma_+ = \sum_{i=0}^{2}\tfrac{1}{3}(|i\rangle|m\rangle\langle i|\langle m|), \qquad (34)$$

$$\sigma_- = \sum_{i=0}^{2}\tfrac{1}{3}(|m\rangle|i\rangle\langle m|\langle i|), \qquad (35)$$

where $m = (i+1) \bmod 3$ [14].

This state contains no distillable entanglement which leads to $Q(\mathcal{N}) = Q_2(\mathcal{N}) = 0$, however it a PPT (Positive Partial Transpose) system, i.e., the partial transposes taken with respect to any subsystems are positive:

$$(\sigma_\alpha)^{T_A} \geq 0 \qquad (36)$$

and

$$(\sigma_\alpha)^{T_{A'}} \geq 0. \qquad (37)$$

The PPT property allows positive private classical capacity $P(\mathcal{N}) \geq 0$ over an entanglement-binding channel $\mathcal{N}$ that generates bound-entangled system $\sigma = (I \otimes \mathcal{N})|\Psi_{AA'}\rangle$, besides the fact that it cannot be used for quantum communication.

## 2.2 Channel Composition

**An anti-degradable quantum channel**

The model of an anti-degradable quantum channel $\mathcal{N}$ is depicted in Fig. 1. The logical channel between Alice and Bob is denoted by $\mathcal{N}_{AB}$, the channel between Alice and the environment is depicted by $\mathcal{N}_{AE}$. For an anti-degradable quantum channel, the channel output $B$ can be expressed from the environment state $E$, using the $\mathcal{D}^{E \to B}$ degradation map. The quantum capacity of an anti-degradable channel is still undetermined. It could be additive or non-additive, depending on the channel model and the dimension of the channel.

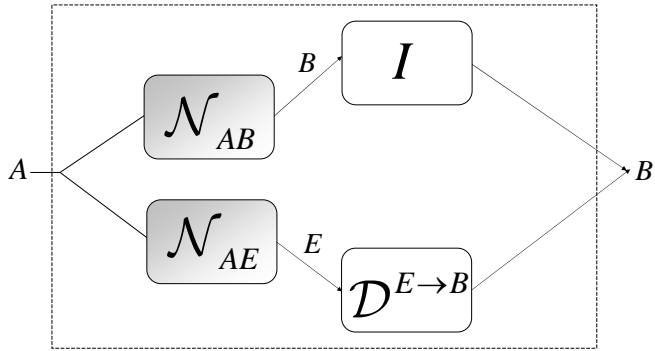

**Figure 1.** The schematic model of an anti-degradable quantum channel $\mathcal{N}$. The logical channel between Alice and Bob is denoted by $\mathcal{N}_{AB}$, the channel between Alice and the environment is depicted by $\mathcal{N}_{AE}$. For an anti-degradable quantum channel, the channel output $B$ can be expressed from the environment state $E$, using the $\mathcal{D}^{E \to B}$ degradation map.

**A degradable quantum channel**

The model of a degradable channel is shown in Fig. 2. The environment state $E$ outputted by $\mathcal{N}_{AE}$ can be expressed from the channel output state outputted by $\mathcal{N}_{AB}$ using the $\mathcal{D}^{B \to E}$ degradation map. The quantum capacity of a degradable channel is provably additive for arbitrary dimensions.

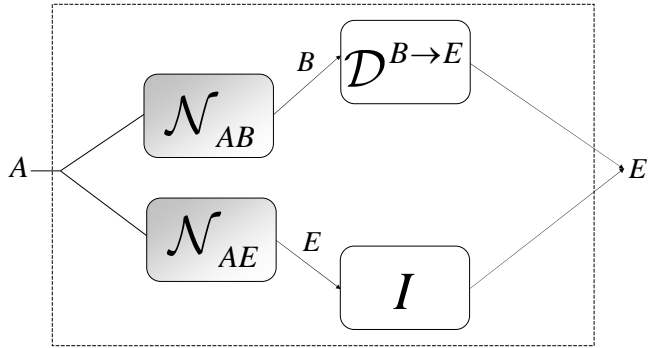

**Figure 2.** The schematic model of a degradable channel. The environment state $E$ outputted by $\mathcal{N}_{AE}$ can be expressed from the channel output state outputted by $\mathcal{N}_{AB}$ using the $\mathcal{D}^{B \to E}$ degradation map.

**A degradable and conjugate degradable quantum channel**

The model of a degradable and conjugate degradable quantum channel is depicted in Fig. 3. The degradation map $\mathcal{D}^{B \to E}$ can be used to simulate $E$ by $B$, but the channel output $B$ also can be used to express $E'$, and from environment state $E$ the complex conjugated environment state $E'$ can be expressed by a complex conjugation $\mathcal{C}$. The quantum capacity of any degradable channel is additive, and it also remains true for a channel that has also the conjugate degradability property. The quantum capacity of a degradable and conjugate degradable channel is always additive in arbitrary dimensions.

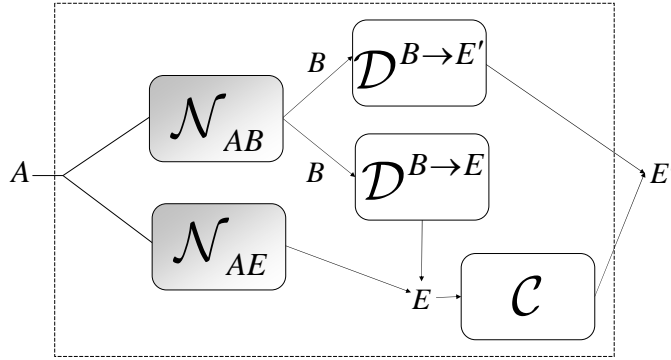

**Figure 3.** The schematic model of a degradable and conjugate degradable quantum channel. The channel output $B$ can be used to express $E'$, and from environment state $E$ the conjugated environment state $E'$ can be expressed up to a complex conjugation $\mathcal{C}$ on $E$.

## 2.3 General Structure of a PD channel

A PD channel is characterized as follows. For a PD channel there exists the partial degradation map $\mathcal{D}^{B \to E'}$ for channel output $B$, from which the degraded environment $E' = \mathcal{N}_{AE} \circ \mathcal{D}^{E \to E'}$ can be simulated, where $\mathcal{N}_{AE}$ is the entanglement-binding complementary channel and $\mathcal{D}^{E \to E'}$ is the degradation map for the environment state $E$. The PD channels can be classified into degradable PD, anti-degradable PD and conjugate-PD sets. Similar to the degradable channels, the quantum capacity of conjugate degradable channels are additive [1]. For a conjugate-PD channel there exist conjugate degradation maps $\mathcal{C}^{E \to E'}$ and $\mathcal{C}^{B \to E'}$ applied on $E$ and $B$, from which a degraded and complex conjugated environment $E'$ can be expressed.

A conjugate-PD channel can be defined both in the degradable (*degradable conjugate-PD*) and anti-degradable (*anti-degradable conjugate-PD*) family, as follows. For a conjugate-PD channel there exists the partial conjugate degradation map $\mathcal{C}^{B \to E'}$ for output $B$, from which the degraded and conjugated environment $E' = \mathcal{N}_{AE} \circ \mathcal{C}^{E \to E'}$ can be simulated, where $\mathcal{N}_{AE}$ is the entanglement-binding complementary channel and $\mathcal{C}^{E \to E'}$ is the conjugate degradation map for environment state $E$. The conjugate-PD channels can be classified into degradable conjugate-PD and anti-degradable conjugate-PD sets. The PD channels are equipped with a well-characterized structure. The complementary channel $\mathcal{N}_{AE}$ of a PD channel is an *entanglement-binding* channel. For any PD channel it will always results in an additive quantum capacity. For any PD channel, the degrading map $\mathcal{D}^{B \to E'}$ on channel output $B$ cannot be used to simulate the environment state $E$, and the channel $\mathcal{N}$ in relation logical channels $\mathcal{N}_{AB}$ and $\mathcal{N}_{AE}$ could be a degradable or anti-degradable one. On the other hand, in relation logical channels $\mathcal{N}_{AB}$ and $\mathcal{N}_{AE'}$, the channel $\mathcal{N}$ is *always degradable*: and that is the property that can be exploited in the evolution of quantum capacity for a PD channel.

**A partially degradable quantum channel**

The general structure of a PD channel is depicted in Fig. 4. The output state $B$ of channel $\mathcal{N}_{AB}$ can be used to simulate the degraded environment state $E'$ by the degradation map $\mathcal{D}^{B \to E'}$, where $\mathcal{N}_{AE'} = \mathcal{N}_{AB} \circ \mathcal{D}^{B \to E'} = \mathcal{N}_{AE} \circ \mathcal{D}^{E \to E'}$. This property is called the *partial simulation* or *partial degradability* property. For any PD channel, the partial simulation makes possible to simulate $E'$ from $B$ by using the partial degrading map $\mathcal{D}^{B \to E'}$.

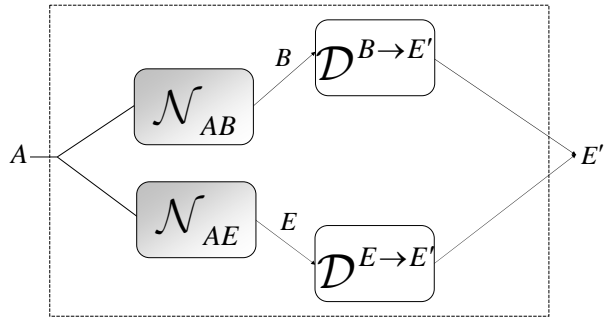

**Figure 4.** The model of a PD quantum channel. For a PD channel, the complementary channel $\mathcal{N}_{AE}$ is an entanglement-binding channel. The degraded environment state $E'$ is outputted by $\mathcal{N}_{AE'} = \mathcal{N}_{AE} \circ \mathcal{D}^{E \to E'}$, where $\mathcal{D}^{E \to E'}$ is an anti-degradable, but non entanglement-breaking map on $E$. For a conjugate-PD channel, the maps $\mathcal{D}^{B \to E'}$ and $\mathcal{D}^{E \to E'}$ represent conjugate degradation that will be denoted by $\mathcal{C}^{B \to E'}$ and $\mathcal{C}^{E \to E'}$, where $\mathcal{C}^{E \to E'}$ is a conjugate anti-degradable map. For a PD channel the output state $B$ of $\mathcal{N}_{AB}$ can be used to simulate $E'$. This property is called partial simulation and has an important consequence as it formulates the quantum capacity of any anti-degradable channel to be additive in arbitrary dimensions.

If both the outputs of $\mathcal{N}_{AE}$ and $\mathcal{N}_{AE'}$ can be simulated by $B$, then the PD channel $\mathcal{N}$ also has the degradability property, and $\mathcal{D}^{B \to E}$ also exists. If the PD channel has also the conjugate degradability property, then using environment state $E$, the conjugate degraded environment $E'$ of $\mathcal{N}_{AE'}$ can be expressed from $B$. It requires a conjugate degrading map $\mathcal{C}^{B \to E'}$, which leads to $\mathcal{N}_{AE'} = \mathcal{N}_{AB} \circ \mathcal{C}^{B \to E'} = \mathcal{N}_{AE} \circ \mathcal{C}^{E \to E'}$.

Next, we briefly overview the composition of the PD family.

**An anti-degradable but conjugate degradable quantum channel**

The model of an anti-degradable but conjugate degradable quantum channel is depicted in Fig. 5. For this channel there is no exists degradation map $\mathcal{D}^{B \to E}$. The channel output $B$ can be used to express $E'$, and from environment state $E$ the complex conjugated environment state $E'$ can be expressed by a complex conjugation $\mathcal{C}$. The quantum capacity of any anti-degradable but conjugate degradable channel is additive.

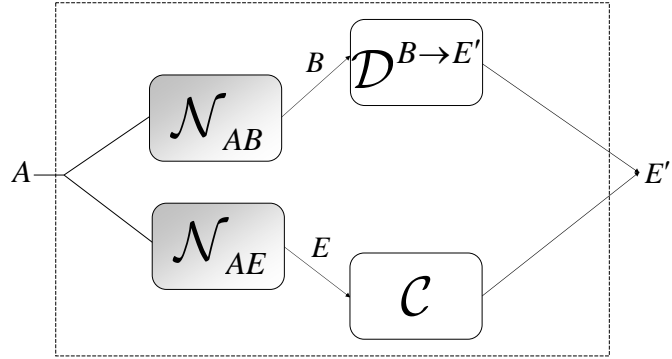

**Figure 5.** The schematic model of an anti-degradable but conjugate degradable quantum channel. The channel output $B$ can be used to express $E'$, and from environment state $E$ the conjugated environment state $E'$ can be expressed by a complex conjugation $\mathcal{C}$ on $E$. There is no exists degradation map $\mathcal{D}^{B \to E}$.

**A conjugate-PD channel**

The model of a conjugate-partially degradable (conjugate-PD) quantum channel is depicted in Fig. 6. The channel output $B$ can be used to express $E'$ using conjugate degradation map $\mathcal{C}^{B \to E'}$, where $E'$ is a complex conjugated and degraded environment. From environment state $E$ the degraded conjugated environment state $E'$ can be expressed by the conjugate degradation map $\mathcal{C}^{E \to E'}$. The quantum capacity of a conjugate-PD channel is additive in arbitrary dimensions.

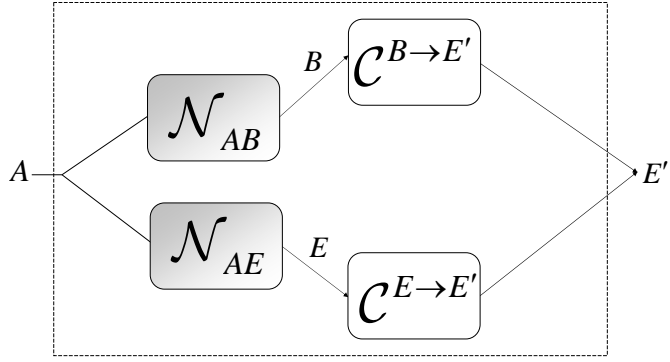

**Figure 6.** The schematic model of a conjugate-PD quantum channel. The channel output $B$ can be used to express the degraded and complex conjugated $E'$, using the conjugated degradation map $\mathcal{C}^{B\to E'}$. From environment state $E$ the degraded environment state $E'$ can be expressed by the conjugate degradation map $\mathcal{C}^{E\to E'}$.

## 3 Theorems and Proofs

The following notations will be used in the detailed proof. The logical channels of quantum channel $\mathcal{N}$ are defined as follows. The logical channels between Alice and Bob and between Alice and the environment are denoted by $\mathcal{N}_{AB}$ and $\mathcal{N}_{AE}$, respectively.

**Notation 1.** *The channel between Alice and Bob, assuming complementary channel $\mathcal{N}_{AE}$, is denoted by $\mathcal{N}_{AB}^{(AE)}$, where $AE$ refers to the complementary channel $\mathcal{N}_{AE}$.*

**Notation 2.** *The channel between Alice and Bob, assuming a degraded complementary channel $\mathcal{N}_{AE'}$, will be referred to as $\mathcal{N}_{AB}^{(AE')}$, where $AE'$ refers to the degraded complementary channel $\mathcal{N}_{AE'} = \mathcal{N}_{AE} \circ \mathcal{D}^{E\to E'}$.*

**Notation 3.** *The degraded channel between Alice and Bob, assuming a degraded complementary channel $\mathcal{N}_{AE'}$, is defined by $\mathcal{N}_{AB'}^{(AE')} = \mathcal{N}_{AB}^{(AE')} \circ \mathcal{D}^{E\to E'}$.*

The input and output dimension of the channels will be denoted as follows. The dimension of the input system of $\mathcal{N}$ is denoted by $d_A$. The output dimensions of the logical channels $\mathcal{N}_{AB}^{(AE)}$, $\mathcal{N}_{AB}^{(AE')}$, $\mathcal{N}_{AB'}^{(AE')}$, and complementary channels $\mathcal{N}_{AE}$ and $\mathcal{N}_{AE'}$ are denoted by $d_B$, $d_E$, and $d_{E'}$. The notations of the proofs are summarized in Fig. 7.

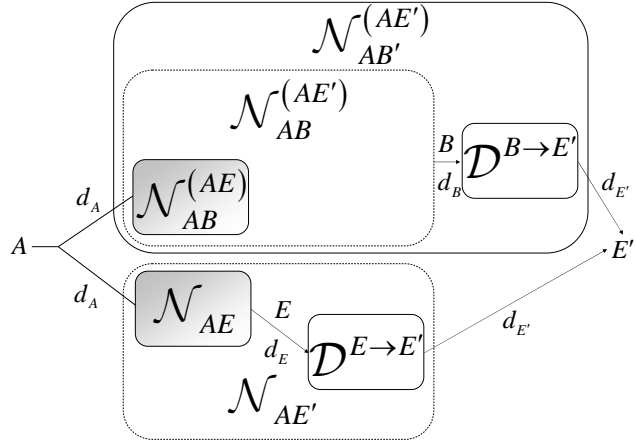

**Figure 7.** The notations of the PD channel structure. An anti-degradable PD channel is anti-degradable but still contains the PD property. The figure depicts the channel input and output systems, the dimensions, and the logical channels.

Our results on the properties of the PD channel structure are summarized in Theorem 1.

**Theorem 1** (On the structure of a PD channel). *For any PD channel $\mathcal{N}$, the complementary channel $\mathcal{N}_{AE}$ is an entanglement-binding channel.*

*Proof.*

Assume that the complementary channel $\mathcal{N}_{AE}$ of $\mathcal{N}$ is degradable, i.e., there exists a degradation map $\mathcal{D}^{E \to E'}$ to simulate $E'$ by $E$. The degradation map $\mathcal{D}^{B \to E'}$ applied on the output $B$ of $\mathcal{N}_{AB}$ can be used only to simulate the output $E'$ of channel $\mathcal{N}_{AE'}$, and the output $E$ of channel $\mathcal{N}_{AE}$ cannot be simulated by this map. For a PD channel, the complementary channel $\mathcal{N}_{AE}$ is *entanglement-binding*, and the degraded environment state $E'$ is a bound-entangled state. From channel output $B$ the bound-entangled $E'$ can be simulated by a degradation map $\mathcal{D}^{B \to E'}$, on the other hand, such degrading map $\mathcal{D}^{B \to E}$ for $B$ could also exists from which $E$ could be simulated but only if the channel $\mathcal{N}$ is degradable. If such a $\mathcal{D}^{B \to E}$ does not exist, then it immediately leads to the conclusion that channel $\mathcal{N}$ is indeed anti-degradable but partially degradable, since only $\mathcal{D}^{B \to E'}$ could exist. This channel is called an anti-degradable PD channel (*Note*: If $\mathcal{C}^{E \to E'}$ conjugate anti-degradable map exists for $\mathcal{N}$ then it is a conjugate-PD channel).

To step further, we use the fact that the channel is PD and a degrading map $\mathcal{D}^{E \to E'}$ for $E$ exists, which makes it possible to simulate $E'$ by $E$.

For channel input $A$, reference system $R$, the following relation holds for a PD channel defined by degradation maps $\mathcal{D}^{E \to E'}$ and $\mathcal{D}^{B \to E'}$:

$$\begin{aligned}
&\left(\mathcal{D}^{E \to E'} \otimes I\right)\left(\mathcal{N}_{AE} \otimes I\right)\rho_{AR} \\
&= \left(\mathcal{N}_{AE} \circ \mathcal{D}^{E \to E'} \otimes I\right)\rho_{AR} \geq 0 \\
&= \left(\mathcal{N}_{AB} \circ \mathcal{D}^{B \to E'} \otimes I\right)\rho_{AR} \geq 0.
\end{aligned} \qquad (38)$$

The degraded environment state $E'$ of channel $\mathcal{N}_{AE'}$ is evaluated by the density matrix $\sigma_{E'R}$ as

$$\begin{aligned}
\sigma_{E'R} &= \left(\mathcal{N}_{AE'} \otimes I\right)\rho_{AR} \\
&= \left(\mathcal{N}_{AE} \circ \mathcal{D}^{E \to E'} \otimes I\right)\rho_{AR},
\end{aligned} \qquad (39)$$

for which the partial transposes are positive:

$$\left(\sigma_{E'R}\right)^{T_{E'}} \geq 0 \text{ and } \left(\sigma_{E'R}\right)^{T_R} \geq 0. \qquad (40)$$

The result on the degraded environment state $E'$ follows from the fact that $\mathcal{N}_{AE}$ is an entanglement-binding complementary channel. For environment state $E$ and density matrix $\sigma_{EP}$, the relation

$$\begin{aligned}
\sigma_{ER} &= \left(\mathcal{N}_{AE} \otimes I\right)\rho_{AR} \\
&= \left(\mathcal{N}_{AE} \otimes I\right)\rho_{AR},
\end{aligned} \qquad (41)$$

along with

$$\left(\sigma_{ER}\right)^{T_E} \geq 0 \text{ and } \left(\sigma_{ER}\right)^{T_R} \geq 0 \qquad (42)$$

also hold.

These positive partial transpose (PPT) states cannot be the result of an *entanglement-breaking* complementary channel, since $\mathcal{N}$ cannot be a Hadamard channel. The entanglement-breaking channel is anti-degradable channel which would lead to contradiction. (*Note*: Any Hadamard channel has entanglement-breaking complementary channel $\mathcal{N}_{AE}$. The Hadamard channel is a degradable channel.) It immediately shows that the output system $\sigma_{E'P}$ of channel $\mathcal{N}_{AE'}$ is a bound-entangled state; in other words, the complementary channel $\mathcal{N}_{AE}$ is an *entanglement-binding* channel. We note that for the complementary channel $\mathcal{N}_{AE}$, for the product of the input and output space dimensions $d_{in}\left(\mathcal{N}_{AE}\right)$ and $d_{out}\left(\mathcal{N}_{AE}\right)$, the condition $d_{in}\left(\mathcal{N}_{AE}\right) \cdot d_{out}\left(\mathcal{N}_{AE}\right) > 6$ has to be satisfied [1], [14,15], otherwise the degrading map $\mathcal{D}^{E \to E'}$ on the output $E$ of the complementary channel $\mathcal{N}_{AE}$ cannot produce bound-entangled system $\sigma_{E'P}$. Since this condition trivially could be satisfied, it does not have any further impact on the proof. From this analysis an important conclusion follows immediately: $\mathcal{N}$ has to be a qudit (dim>2) channel.

For a conjugate-PD channel, the maps $\mathcal{D}^{B \to E'}$ and $\mathcal{D}^{E \to E'}$ are replaced by conjugate degrading maps $\mathcal{C}^{B \to E'}$ and $\mathcal{C}^{E \to E'}$, and the same results and conclusions follow.

∎

For a PD channel $\mathcal{N}$, the output $E'$ of $\mathcal{N}_{AE'} = \mathcal{N}_{AE} \circ \mathcal{D}^{E \to E'}$ can be simulated by the partial degrading map $\mathcal{D}^{B \to E'}$ as $\mathcal{N}_{AB'} = \mathcal{N}_{AB} \circ \mathcal{D}^{B \to E'}$, where $\mathcal{D}^{E \to E'}$ is a degrading map. If degraded environment state $E'$ could not be simulated by $B$, then channel $\mathcal{N}$ indeed would be a degradable or anti-degradable channel, but not a PD channel. In this case $\mathcal{D}^{B \to E'}$ could not exist. This corollary reveals the main idea behind the partial simulation (or partial degradability) property. The channel output of a PD channel with $\mathcal{D}^{B \to E'}$ can be used for only a partial simulation of the environment $E$. For a conjugate-PD channel the same conclusion follows by using the conjugate degrading maps $\mathcal{C}^{B \to E'}$ and $\mathcal{C}^{E \to E'}$.

**Corollary 1** (On the partial simulation property of PD channels). *For a PD channel, there exists a partial degradation map $\mathcal{D}^{B \to E'}$ (for a conjugate-PD channel $\mathcal{C}^{B \to E'}$) that makes it possible to use B to simulate the degraded environment $E'$. The environment state E cannot be simulated by $\mathcal{D}^{B \to E'}$.*

*Proof.*
For a PD channel, the structure of the complementary channel $\mathcal{N}_{AE'}$ makes no possible to use the channel output $B$ for the simulation of the environment $E$. On the other hand, a *partial simulation* of the environment $E$ is possible from channel output $B$. The degraded environment state $E' = \mathcal{N}_{AE} \circ \mathcal{D}^{B \to E'}$ can be simulated by a degrading map $\mathcal{D}^{B \to E'}$, but not the output $E$ of the entanglement-binding complementary channel $\mathcal{N}_{AE}$. The partial simulation property is a benefit that clearly arises from the structure of the $\mathcal{N}_{AE}$ complementary channel of a PD channel. In sense of systems $B$ and $E$ the PD channel $\mathcal{N}$ is anti-degradable, since the environment state $E$ of channel $\mathcal{N}_{AE}$ cannot be simulated by $\mathcal{D}^{B\to E'}$, only the degraded environment state $E' = \mathcal{N}_{AE} \circ \mathcal{D}^{B\to E'}$. Contrary to degradable channels, from the output of the complementary channel $\mathcal{N}_{AE}$ of $\mathcal{N}$ a degraded environment state $E'$ can be constructed by the map $\mathcal{D}^{E\to E'}$. The relation between the logical channels $\mathcal{N}_{AB}$, $\mathcal{N}_{AE}$ and $\mathcal{N}_{AE'}$ of $\mathcal{N}$ also causes a change in the meaning of the degradation.

For an anti-degradable channel there exists no degrading map $\mathcal{D}^{B\to E}$. The partial degrading map $\mathcal{D}^{B\to E'}$ on channel output $B$ is not able to simulate the environment state $E$, and the channel $\mathcal{N}$ in relation logical channels $\mathcal{N}_{AB}$ and $\mathcal{N}_{AE}$ is still anti-degradable, since the environment $E$ cannot be simulated by $B$ and $\mathcal{D}^{B\to E'}$. On the other hand, in sense of $\mathcal{N}_{AB}$ and $\mathcal{N}_{AE'}$, the channel $\mathcal{N}$ is degradable, since $E'$ can be simulated by $B$, using $\mathcal{D}^{B\to E'}$.

For a conjugate-PD channel the same conclusion follows by using the conjugate degrading maps $\mathcal{C}^{B\to E'}$ and $\mathcal{C}^{E\to E'}$.

∎

Since the same results follow for the degradable and anti-degradable PD channels, we omit the proof of Corollaries 2 and 3.

**Corollary 2** (On a degradable PD channel). *If degrading maps $\mathcal{D}^{B\to E}$ and $\mathcal{D}^{B\to E'}$ both exist, then the channel $\mathcal{N}$ is a degradable PD channel.*

**Corollary 3** (On an anti-degradable PD channel). *If degrading map $\mathcal{D}^{B\to E'}$ exists but $\mathcal{D}^{B\to E}$ does not, then $\mathcal{N}$ is an anti-degradable PD channel.*

For a conjugate-PD channel the same corollaries follow by using the conjugate degrading maps $\mathcal{C}^{B\to E'}$ and $\mathcal{C}^{E\to E'}$.

The main result for the behavior of quantum capacity of the defined PD channel structure is summarized in Theorem 2.

**Theorem 2** (On the quantum capacity of PD channels). *The quantum capacity of a PD and conjugate-PD channel is additive.*

*Proof.*

The quantum capacity of the PD channel is evaluated from the outputs $B$, $E$, and $E'$ of the channels $\mathcal{N}_{AB}$, $\mathcal{N}_{AE}$, and $\mathcal{N}_{AE'}$, and by the degrading maps $\mathcal{D}^{B \to E'}$ and $\mathcal{D}^{E \to E'}$. To describe the maps by the Stinespring dilation, we also add the auxiliary systems $F$ and $H$. For a PD channel, the map of channel $\mathcal{N}_{AB}$ and the degradation maps $\mathcal{D}^{E \to E'}$ and $\mathcal{D}^{B \to E'}$ can be represented by the isometries $U: A \to BE$, $V: E \to GH$ and $W: B \to E'F$. The dimension of the $i$-th system is denoted by $d_i$.

The isometries of a PD channel are depicted in Fig. 8. The input is denoted by $A$, the system formulated by the input state $A$ and the reference state $R$ is depicted by $|\phi\rangle_{AR}$. The state shared between systems $G$, $H$, $B$ and $R$ is referred by $|\varphi\rangle$. The state shared between $G$, $H$, $E'$, $F$ and $R$ is denoted by $|\psi\rangle$.

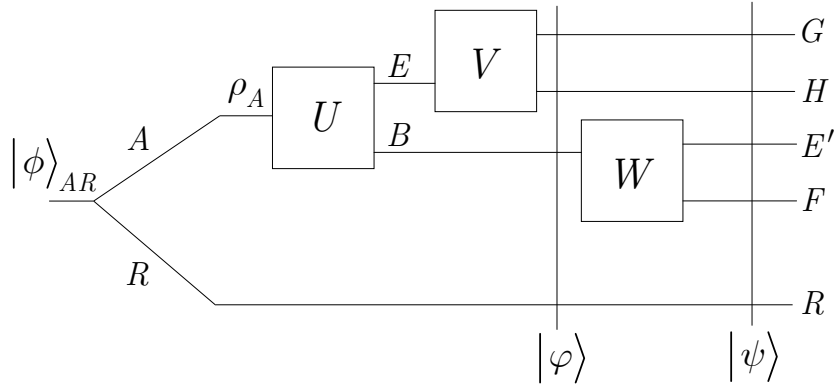

**Figure 8.** The isometries of a PD channel.

Assuming input $\rho_A$ and environment state $\rho_E$, the output $\tau_B$, degraded environment $\tau_{E'}$ and $\omega_{E'}$ of logical channels $\mathcal{N}_{AB}$ and $\mathcal{N}_{AE'}$ are defined as

$$\tau_B = U\rho_A U^\dagger, \tag{43}$$

$$\tau_{E'} = V\rho_E V^\dagger, \tag{44}$$

$$\omega_{E'} = W\tau_B W^\dagger. \tag{45}$$

For a PD channel the following relation holds for logical channels $\mathcal{N}_{AB'} = \mathcal{N}_{AB} \circ \mathcal{D}^{B \to E'}$ and $\mathcal{N}_{AE'} = \mathcal{N}_{AE} \circ \mathcal{D}^{E \to E'}$:

$$\mathcal{N}_{AB'} = \mathcal{N}_{AE'} \tag{46}$$
$$\mathcal{N}_{AB} \circ \mathcal{D}^{B \to E'} = \mathcal{N}_{AE} \circ \mathcal{D}^{E \to E'},$$

which is equal to the following statement:

$$\psi_{GR} = \psi_{E'R}, \tag{47}$$

modulo the isomorphism between $G$ and $E'$. The vector states $\psi_{GH}$ and $\psi_{FE'}$ are linear functions of the reference state $\rho_R = Tr_A(\rho_{AR})$.

The coherent information of a PD channel can be expressed as follows:

$$I_{coh}(\mathcal{N}, \rho_A)_\varphi = H(F|E')_\psi = H(H|G)_\psi, \tag{48}$$

where $H(\cdot)$ is the von Neumann entropy function.

To show it we have to prove that $\psi_{GR} = \psi_{E'R}$. First we define two unitaries $U_H$ and $U_F$ on systems $H$ and $F$, and then two isometries $Y$ and $Z$ [16]. Let denote by $U^{swap}$ the unitary swap operation, defined as $U^{swap}|a\rangle|b\rangle = |b\rangle|a\rangle$. The equation $\psi_{GR} = \psi_{E'R}$ could hold if only if

$$U^{swap}|G\rangle|E'\rangle = |E'\rangle|G\rangle \tag{49}$$

is also satisfied modulo the implicit identification of $G$ with $E'$. To see it we define auxiliary unitary operators on systems $H$ and $F$ as follows [16]:

$$U_H : U_H^2 = I_{d_H}, \tag{50}$$

and

$$U_F : U_F^2 = I_{d_F}, \tag{51}$$

where $I_{d_H}$ and $I_{d_F}$ are the $d_H$ and $d_F$ dimensional identity matrices.

To demonstrate that $E' = G$, we show that Stinespring dilations $Y : VU$ and $Z : WU$ are, in fact, equivalent. These isometries are defined as

$$Y : VU : A \to GE'HI \tag{52}$$

and

$$Z : WU : A \to GE'FJ, \tag{53}$$

where auxiliary systems $I$ and $J$ have dimensions $d_I = 2$ and $d_J = 2$, respectively.

Using the auxiliary unitary operators $U_H$, $U_F$ and $U^{swap}$ on environment states $G$ and $E'$, along with the Stinespring dilations $Y : VU$ and $Z : WU$, where $U : A \to BE$, $V : E \to GH$ and $W : B \to E'F$ of maps $\mathcal{N}_{AB}$, $\mathcal{D}^{E \to E'}$ and $\mathcal{D}^{B \to E'}$ lead to the following:

$$\left(U_H \otimes U^{swap}|E'\rangle|G\rangle\right)Y = Y \tag{54}$$

and
$$\left(U_F \otimes U^{swap}\left|G\right\rangle\left|E'\right\rangle\right)Z = Z. \tag{55}$$

Identifying input system by state vector $\left|\phi\right\rangle_{AR}$ and the channel output by the state vectors $\left|\psi\right\rangle_{GE'HR}$ and $\left|\psi\right\rangle_{GE'FR}$, where

$$\left|\psi\right\rangle_{GE'HR} = \left(I_{d_H} \otimes Y\right)\left|\phi\right\rangle_{AR}, \tag{56}$$

and

$$\left|\psi\right\rangle_{GE'FR} = \left(I_{d_F} \otimes Z\right)\left|\phi\right\rangle_{AR}. \tag{57}$$

Applying $U^{swap}$ we get that

$$\begin{aligned}&\left(I_{d_H} \otimes U_H \otimes U^{swap}\left|E'\right\rangle\left|G\right\rangle\right)\left|\psi\right\rangle_{GE'HR} \\ &= \left(I_{d_H} \otimes U_H \otimes U^{swap}\left|E'\right\rangle\left|G\right\rangle\right)\left(I_{d_H} \otimes Y\right)\left|\phi\right\rangle_{AR} \\ &= \left|\psi\right\rangle_{GE'HR},\end{aligned} \tag{58}$$

and

$$\begin{aligned}&\left(I_{d_F} \otimes U_F \otimes U^{swap}\left|G\right\rangle\left|E'\right\rangle\right)\left|\psi\right\rangle_{GE'FR} \\ &= \left(I_{d_F} \otimes U_F \otimes U^{swap}\left|G\right\rangle\left|E'\right\rangle\right)\left(I_{d_F} \otimes Z\right)\left|\phi\right\rangle_{AR} \\ &= \left|\psi\right\rangle_{GE'FR}.\end{aligned} \tag{59}$$

The isometries $Y$ and $Z$ can be rewritten as

$$Y = \frac{1}{\sqrt{2}}\left(V_0 U \otimes \left|0\right\rangle_I + U^{swap}\left|E'\right\rangle\left|G\right\rangle V_0 U \otimes \left|1\right\rangle_I\right), \tag{60}$$

where dilation $V_0$ is defined as

$$V_0 : E \to G_0 H \tag{61}$$

and

$$Z = \frac{1}{\sqrt{2}}\left(W_0 U \otimes \left|0\right\rangle_J + U^{swap}\left|G\right\rangle\left|E'\right\rangle W_0 U \otimes \left|1\right\rangle_J\right), \tag{62}$$

where $W_0$ is defined as

$$W_0 : B \to E'_0 F \tag{63}$$

and systems $I$ and $J$ have dimensions $d_I = 2$ and $d_J = 2$. Rewriting systems $H$ and $F$ as

$$H = H_0 \otimes E' \tag{64}$$

and

$$F = F_0 \otimes G, \tag{65}$$

along with unitary operators

$$U_H = I_{d_{H_0}} \otimes X, \tag{66}$$

$$U_F = I_{d_{F_0}} \otimes X, \tag{67}$$

where $X$ is the Pauli $X$ operator, we get

$$Y = \left(U^{swap}\left|E'\right\rangle\left|G\right\rangle \otimes X\right)Y \tag{68}$$

and

$$Z = \left(U^{swap}\left|G\right\rangle\left|E'\right\rangle \otimes X\right)Z. \tag{69}$$

To see that systems $E' = G$ are equivalent we extend the Stinespring dilations $Y$ and $Z$ as follows. For isometry $Y$ we have

$$\begin{aligned} Tr_{GH} Y\rho Y^\dagger &= \mathcal{N}_{AB'}(\rho) = \mathcal{N}_{AB} \circ \mathcal{D}^{B \to E'} \\ &= W\tau_B W^\dagger \\ &= \omega_{E'}, \end{aligned} \tag{70}$$

where system $\tau_B$ was defined in (43) and $W$ is the isometry $W : B \to E'F$. For isometry $Z$ we obtain

$$\begin{aligned} Tr_{E'F} Z\rho Z^\dagger &= \mathcal{N}_{AE'}(\rho) = \mathcal{N}_{AE} \circ \mathcal{D}^{E \to E'} \\ &= V\rho_E V^\dagger \\ &= \tau_{E'}, \end{aligned} \tag{71}$$

where $\rho_E$ is the environment state outputted by complementary channel $\mathcal{N}_{AE}$, and $V$ is the Stinespring dilation $V : E \to GH$. As follows, the Stinespring dilations $Y : VU$ and $Z : WU$ are equivalent and isometries

$$V : E \to GH \tag{72}$$

and

$$W : B \to E'F \tag{73}$$

provably exist for a PD channel. Isometry $W$ can be used to simulate the degraded environment state $G$ from $B$ such that it results in $E' = G$, which can be demonstrated by $Y$ and $Z$ as

$$\begin{aligned} Y &= \left(U^{swap}\left|E'\right\rangle\left|G\right\rangle \otimes U_H\right)Y \\ VU &= \left(U^{swap}\left|E'\right\rangle\left|G\right\rangle \otimes U_H\right)VU, \end{aligned} \tag{74}$$

and

$$\begin{aligned} Z &= \left(U^{swap}\left|G\right\rangle\left|E'\right\rangle \otimes U_F\right)Z \\ WU &= \left(U^{swap}\left|G\right\rangle\left|E'\right\rangle \otimes U_F\right)WU. \end{aligned} \tag{75}$$

These results clearly show that for a PD channel systems $E'$ and $G$ are equivalent, i.e., our conclusion is

$$E' = G. \tag{76}$$

From these results follows that for a PD channel $\mathcal{N}$, there exist degrading maps $\mathcal{D}^{E \to E'}$ and $\mathcal{D}^{B \to E'}$ represented by Stinespring dilations $V : E \to GH$ and $W : B \to E'F$. In other words, for a PD channel the coherent information can be evaluated by systems $B$ and $E'$, where $E' = G$. For the channel output $\tau_B$, and the degraded environment $\omega_{E'}$ the relation $H(B)_{\tau_B} = H(E'F)_{\omega_{E'}}$ holds, and the quantum coherent information of $\mathcal{N}$ for input system $\rho$ can be evaluated as

$$I_{coh}(\mathcal{N}, \rho) = H(B)_{\tau_B} - H(E')_{\tau_B} = H(E'F)_{\omega_{E'}} - H(E')_{\omega_{E'}} = H(F|E')_{\omega_{E'}}. \tag{77}$$

(*Note*: One can readily conclude, that for the standard (non-PD) degraded case $H(E')_{\tau_B} = H(E)_{\tau_B}$, which is trivially not the case in (77)). Now let assume that we use the channel $\mathcal{N}$ for $n$ times, denoted by $\mathcal{N}^{\otimes n}$. The channel $\mathcal{N}_{AB}$ is defined with $n$ inputs $A_1, \ldots, A_n$ and output systems $B_1, \ldots, B_n$, the complementary channel $\mathcal{N}_{AE}$ outputs $E_1, \ldots, E_n$. The output of the degraded complementary channel $\mathcal{N}_{AE'}$ is $E'_1, \ldots, E'_n$. Using the Stinespring dilation, the auxiliary system is expressed as $F_1, \ldots, F_n$.

Using the isometries $U : A \to BE$, $V : E \to GH$ and $W : B \to E'F$ of a PD channel, for any $n$-length input $\rho_{A_1, \ldots, A_n}$, the outputs of logical channels $\mathcal{N}_{AB}^{\otimes n}$, $\mathcal{N}_{AE}^{\otimes n}$ and $\mathcal{N}_{AE'}^{\otimes n}$ of $\mathcal{N}^{\otimes n}$ are expressed as

$$\tau_{B_1, \ldots, B_n} = (U_1 \otimes \ldots \otimes U_n) \rho_{A_1, \ldots, A_n} (U_1^\dagger \otimes \ldots \otimes U_n^\dagger), \tag{78}$$

$$\tau_{E'_1, \ldots, E'_n} = (V_1 \otimes \ldots \otimes V_n) \rho_{E_1, \ldots, E_n} (V_1^\dagger \otimes \ldots \otimes V_n^\dagger) \tag{79}$$

and

$$\begin{aligned}\omega_{E'_1, \ldots, E'_n} &= (W_1 \otimes \ldots \otimes W_n)\left((U_1 \otimes \ldots \otimes U_n) \rho_{A_1, \ldots, A_n} (U_1^\dagger \otimes \ldots \otimes U_n^\dagger)\right)(W_1^\dagger \otimes \ldots \otimes W_n^\dagger) \\ &= (W_1 \otimes \ldots \otimes W_n) \tau_{B_1, \ldots, B_n} (W_1^\dagger \otimes \ldots \otimes W_n^\dagger).\end{aligned} \tag{80}$$

For the output system of $\mathcal{N}^{\otimes n}$, the strong subadditivity of von Neumann entropy function leads to the following inequality:

$$\begin{aligned}H(F_1 \ldots F_n | E'_1 \ldots E'_n)_{\omega_{E'_1, \ldots, E'_n}} &\leq \\ H(F_1 | E'_1 \ldots E'_n)_{\omega_{E'_1, \ldots, E'_n}} + H(F_2 | E'_1 \ldots E'_n)_{\omega_{E'_1, \ldots, E'_n}} &+ \ldots + H(F_n | E'_1 \ldots E'_n)_{\omega_{E'_1, \ldots, E'_n}} \leq \\ H(F_1 | E'_1)_{\omega_{E'_1, \ldots, E'_n}} + H(F_2 | E'_2)_{\omega_{E'_1, \ldots, E'_n}} &+ \ldots + H(F_n | E'_n)_{\omega_{E'_1, \ldots, E'_n}}.\end{aligned} \tag{81}$$

The coherent information of channel $\mathcal{N}^{\otimes n}$ can be expressed as follows:

$$\begin{aligned}
&I_{coh}\left(\mathcal{N}^{\otimes n}, \rho_{A_1,\ldots,A_n}\right) \\
&= H\left(B_1,\ldots,B_n\right)_{\tau_{B_1,\ldots,B_n}} - H\left(E'_1,\ldots,E'_n\right)_{\tau_{B_1,\ldots,B_n}} \\
&= H\left(E'_1,\ldots,E'_n F_1,\ldots,F_n\right)_{\omega_{E'_1,\ldots,E'_n}} - H\left(E'_1,\ldots,E'_n\right)_{\omega_{E'_1,\ldots,E'_n}} \\
&= H\left(F_1,\ldots,F_n \,\middle|\, E'_1,\ldots,E'_n\right)_{\omega_{E'_1,\ldots,E'_n}},
\end{aligned} \tag{82}$$

where

$$I_{coh}\left(\mathcal{N}^{\otimes n}, \rho_{A_1,\ldots,A_n}\right) \leq I_{coh}\left(\mathcal{N}, \rho_{A_1}\right) + I_{coh}\left(\mathcal{N}, \rho_{A_2}\right) + \ldots + I_{coh}\left(\mathcal{N}, \rho_{A_n}\right). \tag{83}$$

From the coherent information of the channel $\mathcal{N}^{\otimes n}$, for the quantum capacity of the channel, this conclusion follows:

$$Q\left(\mathcal{N}^{\otimes n}\right) \leq n \cdot Q^{(1)}\left(\mathcal{N}\right), \tag{84}$$

hence, the quantum capacity single-letterizes:

$$Q\left(\mathcal{N}\right) = Q^{(1)}\left(\mathcal{N}\right) = \max_{\forall \rho_A} I_{coh}\left(\mathcal{N}, \rho_A\right). \tag{85}$$

The existence of the degrading maps $\mathcal{D}^{B \to E'}$ and $\mathcal{D}^{E \to E'}$ lead to additive quantum capacity for a PD quantum channel in arbitrary dimensions. The PD property can be exploited for the evolution of the quantum capacity of an anti-degradable PD channel.

The coherent information of a PD channel obtained in (85) also can be evaluated on the pure system *BCR*. The chain rule identity of the conditional entropy function leads to

$$H\left(RB \,\middle|\, C\right) = H\left(B \,\middle|\, C\right) + H\left(R \,\middle|\, BC\right). \tag{86}$$

Exploiting the invariance and the duality identity of conditional entropy function we get

$$H\left(R \,\middle|\, B\right) = H\left(RB\right) - H\left(B\right), \tag{87}$$

$$H\left(R \,\middle|\, B\right) = -H\left(R \,\middle|\, C\right), \tag{88}$$

i.e., the previously derived result in (82) can be rewritten as follows

$$\begin{aligned}
&I_{coh}\left(\mathcal{N}^{\otimes n}, \rho_{A_1,\ldots,A_n}\right) \\
&= -H\left(R \,\middle|\, B\right)_\varphi \\
&= -H\left(R \,\middle|\, FE'\right)_\psi \\
&= H\left(F \,\middle|\, E'\right) - H\left(RF \,\middle|\, E'\right) \\
&= H\left(F \,\middle|\, E'\right) + H\left(RF \,\middle|\, E'\right),
\end{aligned} \tag{89}$$

which is equivalent to

$$\begin{aligned}
&I_{coh}\left(\mathcal{N}^{\otimes n}, \rho_{A_1,\ldots,A_n}\right) \\
&= H\left(F \,\middle|\, E'\right) + H\left(RF \,\middle|\, G\right),
\end{aligned} \tag{90}$$

where

$$H(RF|G) = H(RH|G) = H(RH|E') = H(RF|E') = 0. \tag{91}$$

These lead to coherent information

$$I_{coh}\left(\mathcal{N}^{\otimes n}, \rho_{A_1,\ldots,A_n}\right) = H(F|E') = H(H|G). \tag{92}$$

From these results for a quantum capacity of a PD channel follows that

$$Q(\mathcal{N}) = Q^{(1)}(\mathcal{N}) = \max_{\forall \rho_A} H(H|G) = \max_{\forall \rho_A} H(F|E'), \tag{93}$$

which concludes the proof.

The results trivially follows for a conjugate-PD channel, by replacing degrading maps $\mathcal{D}^{B \to E'}$ and $\mathcal{D}^{E \to E'}$ by the conjugate degradation maps $\mathcal{C}^{B \to E'}$ and $\mathcal{C}^{E \to E'}$

∎

As we reveal in the next theorem, arbitrary dimensional anti-degradable PD channels can be constructed by arbitrary dimensional anti-degradable qudit channels that are equipped with an entanglement-binding complementary channel by applying an arbitrary anti-degradable (or conjugate anti-degradable) but non entanglement-breaking noisy CPTP degrading map (i.e., non-identity map) on the complementary channel.

**Theorem 3** (On the construction of an anti-degradable PD channel). *Applying a noisy anti-degradable, non entanglement-breaking CPTP map $\mathcal{D}^{E \to E'}$ (or conjugate anti-degradable map $\mathcal{C}^{E \to E'}$) on the entanglement-binding complementary channel $\mathcal{N}_{AE}$ of any anti-degradable qudit channel $\mathcal{N}$, results in an anti-degradable PD channel only if $\mathcal{D}^{B \to E'} = \left(\mathcal{N}_{AB}^{AE'}\right)^{-1} \circ \mathcal{N}_{AE} \circ \mathcal{D}^{E \to E'}$, such that $\mathcal{N}_{AB}^{AE'} \circ \mathcal{D}^{B \to E'} = \mathcal{N}_{AE} \circ \mathcal{D}^{E \to E'}$. For a conjugate degradable channel $\mathcal{D}^{B \to E'} = \left(\mathcal{N}_{AB}^{AE'}\right)^{-1} \circ \mathcal{N}_{AE} \circ \mathcal{C}^{E \to E'}$ such that $\mathcal{N}_{AB}^{AE'} \circ \mathcal{D}^{B \to E'} = \mathcal{N}_{AE} \circ \mathcal{C}^{E \to E'}$.*

*Proof.*

We demonstrate the results with the maps of $\mathcal{D}^{B \to E'}$ and $\mathcal{D}^{E \to E'}$. First let us define the required conditions on the existence of degrading map $\mathcal{D}^{B \to E'}$. Assuming channel $\mathcal{N} : M_{d_A} \mapsto M_{d_B}$ with $d_B \leq d_A$ and logical channels $\mathcal{N}_{AB}^{AE} : M_{d_A} \mapsto M_{d_B}$, $\mathcal{N}_{AE} : M_{d_A} \mapsto M_{d_E}$, the degradation map $\mathcal{D}^{B \to E'}$ is a CPTP map

$$\mathcal{D}^{B \to E'} : M_{d_B} \mapsto M_{d_{E'}}, \tag{94}$$

as

$$\begin{aligned} \mathcal{D}^{B \to E'} &= \left( \mathcal{N}_{AB}^{AE'} \right)^{-1} \circ \mathcal{N}_{AE'} \\ \mathcal{D}^{B \to E'} &= \left( \mathcal{N}_{AB}^{AE'} \right)^{-1} \circ \mathcal{N}_{AE} \circ \mathcal{D}^{E \to E'}, \end{aligned} \quad (95)$$

on $\left[ \ker \mathcal{N}_{AB}^{AE'} \right]^{\perp}$, where $\mathcal{N}_{AE'}$ is the degraded complementary channel $\mathcal{N}_{AE'} : M_{d_A} \mapsto M_{d_{E'}}$, and $\mathcal{N}_{AB}^{AE'} : M_{d_A} \mapsto M_{d_B}$ is the same channel as $\mathcal{N}_{AB}^{AE} : M_{d_A} \mapsto M_{d_B}$ assuming the degraded complementary channel $\mathcal{N}_{AE'}$.

The map $\mathcal{D}^{E \to E'}$ has to be an anti-degradable map; otherwise, there would exist another map $\eta^{E' \to B}$ given by

$$\eta^{E' \to B} = \left( \mathcal{N}_{AE'} \right)^{-1} \circ \mathcal{N}_{AB}^{AE'}, \quad (96)$$

that could be used to simulate $B$ from $E'$, which would lead to a *symmetric* PD channel as follows:

$$\begin{aligned} \mathcal{N}_{AE'} \circ \eta^{E' \to B} &= \mathcal{N}_{AE} \circ \mathcal{D}^{E \to E'} \circ \eta^{E' \to B} \\ &= \mathcal{N}_{AE} \circ \mathcal{D}^{E \to E'} \circ \left( \mathcal{N}_{AE'} \right)^{-1} \circ \mathcal{N}_{AB}^{AE'} \\ &= \mathcal{N}_{AE'} \circ \left( \mathcal{N}_{AE'} \right)^{-1} \circ \mathcal{N}_{AB}^{AE'} \\ &= \mathcal{N}_{AB}^{AE'}, \end{aligned} \quad (97)$$

which is a contradiction. Therefore, $\mathcal{D}^{E \to E'}$ cannot be a degradable map.

Applying an anti-degradable noisy CPTP degrading map $\mathcal{D}^{E \to E'}$ on the entanglement-binding complementary channel $\mathcal{N}_{AE}$ results in channel

$$\mathcal{N}_{AE'} = \mathcal{N}_{AE} \circ \mathcal{D}^{E \to E'}, \quad (98)$$

which cannot be an entanglement-breaking channel due to the assumptions on $\mathcal{D}^{E \to E'}$. If degrading map $\mathcal{D}^{E \to E'}$ would be an $I$ ideal map, then it would not cause any change in the complementary channel, which would lead to

$$\mathcal{N}_{AE'} = \mathcal{N}_{AE} \circ I = \mathcal{N}_{AE}, \quad (99)$$

and

$$\mathcal{N}_{AB}^{AE} = \mathcal{N}_{AB}^{AE'}, \quad (100)$$

which clearly shows that $\mathcal{N}_{AB}^{AE'}$ would remain still anti-degradable.

On the other hand, if $\mathcal{D}^{E \to E'}$ were an *entanglement-breaking* map, $\Gamma$, then it would lead to an entanglement-breaking complementary channel

$$\mathcal{N}_{AE'} = \mathcal{N}_{AE} \circ \Gamma = \Gamma_{AE'}, \quad (101)$$

which would be equivalent to a Hadamard channel $\mathcal{N}$, which is a contradiction (*Note*: the entanglement-breaking map is also an anti-degradable map). A PD channel cannot have an entanglement-breaking complementary channel $\Gamma_{AE'}$. The action of the entanglement-breaking complementary channel $\Gamma_{AE'}$ on the channel input system $\rho_{AA'}$ could be expressed as

$$\Gamma_{AE'}\left(\rho_{AA'}\right) = I\left(\rho_A\right) \otimes \Gamma_{AE'}\left(\rho_{A'}\right) = \sum_i N_i^{(A')} \rho_{AA'} N_i^{(A')\dagger}, \tag{102}$$

where

$$N_i^{(A')} = I_A \otimes \left|\xi_i\right\rangle_{E'} \left\langle \varsigma\right|_{A'}, \tag{103}$$

where $A'$ and $E'$ denote the input and environment, and $N_i^{(A')}$ are the Kraus-operators with relation $\sum_i N_i^{(A')\dagger} N_i^{(A')} = I$, where each $N_i^{(A')}$ has rank of one. The sets $\left\{\left|\xi_i\right\rangle_{E'}\right\}$ and $\left\{\left|\varsigma\right\rangle_{A'}\right\}$ each do not necessarily form an orthonormal set.

∎

**Corollary 4** *For any PD channel, the map $\mathcal{D}^{E \to E'}$ (or conjugate anti-degradable map $\mathcal{C}^{E \to E'}$) and the degraded complementary channel $\mathcal{N}_{AE'} = \mathcal{N}_{AE} \circ \mathcal{D}^{E \to E'}$, $\mathcal{N}_{AE'} = \mathcal{N}_{AE} \circ \mathcal{C}^{E \to E'}$ cannot be expressed as the sum of rank-one Kraus operators.*

*Proof.*

We demonstrate the results with the map $\mathcal{D}^{E \to E'}$. Let assume that

$$\mathcal{N}_{AE'} = \mathcal{N}_{AE} \circ \mathcal{D}^{E \to E'} = \sum_i N_i^{(A')\dagger} N_i^{(A')} = I, \tag{104}$$

where each $N_i^{(A')}$ are rank-one Kraus operators. The equation in (104) could hold if and only if the degrading map $\mathcal{D}^{E \to E'}$ were entanglement-breaking

$$\mathcal{D}^{E \to E'} = \Gamma, \tag{105}$$

since $\mathcal{N}_{AE}$ is an entanglement-binding channel. The resulting degraded complementary channel is $\mathcal{N}_{AE'} = \Gamma_{AE'}$, which constitutes an entanglement-breaking map.

Let $\mathcal{N}_{AE}$ be the $d_A = d_E = 3$ dimensional entanglement-binding channel from (30) as

$$\sigma_{E'} = \mathcal{N}_{AE}\left(\rho_{A'}\right) = \tfrac{2}{7}\left(\rho_{A'}\right) + \tfrac{\alpha}{7} \sum_{k=1}^{3} \left|j\right\rangle\left\langle k\right|\left(\rho_{A'}\right)\left|k\right\rangle\left\langle j\right| + \tfrac{5-\alpha}{7} \sum_{k=1}^{3} \left|l\right\rangle\left\langle k\right|\left(\rho_{A'}\right)\left|k\right\rangle\left\langle l\right|, \tag{106}$$

where $|\Psi_{AA'}\rangle$ is the $d_A = 3$ dimensional maximally entangled input system $|\Psi_{AA'}\rangle = \frac{1}{\sqrt{3}}\sum_{i=0}^{2}|i\rangle|i\rangle$, $3 < \alpha \leq 4$, $j = (k+1) \mod 3$, and $l = (k-1) \mod 3$.

Let us choose a three-dimensional degrading map $\mathcal{D}^{E \to E'}$ as follows:

$$\mathcal{D}^{E \to E'} : M_{d_E} \mapsto M_{d_{E'}} : M_3 \mapsto M_3, \tag{107}$$

$$\sigma_E' = \mathcal{D}^{E \to E'}(\sigma_E) \equiv \sum_i N_i^{(E)} \sigma_E \left(N_i^{(E)}\right)^\dagger, \tag{108}$$

where $N_i^{(E)}$ are the set of Kraus operators of the degrading map:

$$\begin{aligned} N_i^{(E)} &= \tfrac{1}{\sqrt{4}}|1\rangle_{E'}\langle 1|_E, \tfrac{1}{\sqrt{4}}|2\rangle_{E'}\langle 2|_E, \tfrac{1}{\sqrt{4}}|3\rangle_{E'}\langle 3|_E, \\ &= \tfrac{1}{\sqrt{64}}|\gamma\rangle_{E'}\langle\gamma|_E|\kappa\rangle_{E'}\langle\kappa|_E, \end{aligned} \tag{109}$$

where

$$|\gamma\rangle_{E'} = \sum_{j=1}^{3} i^{n_j}|j\rangle_{E'}, \langle\gamma|_E = \sum_{j=1}^{3} i^{n_j}\langle j|_E, \tag{110}$$

$$|\kappa\rangle_{E'} = \sum_{j=1}^{3} \sqrt{(-1)^{n_j}}|j\rangle_{E'}, \langle\kappa|_E = \sum_{j=1}^{3} \sqrt{(-1)^{n_j}}\langle j|_E, \tag{111}$$

and $\vec{n} = \{n_1, n_2, n_3\}$ with $n_1 = 0$, $n_2, n_3 \in \{0, 1, 2, 3\}$.

Our goal is to show that the degraded complementary channel $\mathcal{N}_{AE'} = \mathcal{N}_{AE} \circ \mathcal{D}^{E \to E'}$ had become entanglement-breaking and $\sum_i N_i^{(A')\dagger} N_i^{(A')} = I$, where $N_i^{(A')}$ are rank-one Kraus operators.

The output system

$$\begin{aligned} \sigma_{AE'} &= \left(I \otimes \mathcal{N}_{AE}(\rho_{A'}) \circ \mathcal{D}^{E \to E'}(\sigma_E)\right)|\Psi_{AA'}\rangle \\ &= \left(I \otimes \mathcal{N}_{AE'}(\rho_{A'})\right)|\Psi_{AA'}\rangle \end{aligned} \tag{112}$$

is a separable system and can be rewritten as

$$\begin{aligned} \sigma_{AE'} &= \sum_{i=1}^{3}\sum_{j=1}^{3}|i\rangle\langle j| \otimes \mathcal{N}_{AE'}(|i\rangle\langle j|) \\ &= \sum_l c_l \varsigma_l \otimes \omega_l \\ &= \sum_l b_l \varsigma_l \otimes \tfrac{1}{3}\left(4I_3 + \sum \overline{c_l} G_l\right), \end{aligned} \tag{113}$$

where $b_l$ and $c_l$ are coefficients, $\overline{c_l}$ is the complex conjugate of coefficient $c_l$, and $\varsigma_l$ and $\omega_l$ are density matrices, while $G_l$ is a generator of the symmetric representation of the $sl(3,\mathbb{C})$ algebra [2],[17-19]. The Kraus operators of channel $\mathcal{N}_{AE'}$ are rank-one Kraus matrices as follows:

$$N_0^{(A')} = \tfrac{1}{\sqrt{4}} |1\rangle_{E'} \langle 1|_{A'},$$
$$N_1^{(A')} = \tfrac{1}{\sqrt{4}} |2\rangle_{E'} \langle 2|_{A'},$$
$$N_2^{(A')} = \tfrac{1}{\sqrt{4}} |3\rangle_{E'} \langle 3|_{A'},$$
$$N_3^{(A')} = \tfrac{1}{\sqrt{64}} |\Upsilon\rangle_{1,E'} \langle \Upsilon|_{1,A'} |\vartheta\rangle_{1,E'} \langle \vartheta|_{1,A'},$$
$$N_4^{(A')} = \tfrac{1}{\sqrt{64}} |\Upsilon\rangle_{2,E'} \langle \Upsilon|_{2,A'} |\vartheta\rangle_{2,E'} \langle \vartheta|_{2,A'},$$
$$N_5^{(A')} = \tfrac{1}{\sqrt{64}} |\Upsilon\rangle_{3,E'} \langle \Upsilon|_{3,A'} |\vartheta\rangle_{3,E'} \langle \vartheta|_{3,A'}, \quad (114)$$

where

$$|\Upsilon\rangle_{j,E'} = i^{n_j} |j\rangle_{E'}, \quad \langle \Upsilon|_{j,A'} = i^{n_j} \langle j|_{A'}, \quad (115)$$

$$|\vartheta\rangle_{j,E'} = \sqrt{(-1)^{n_j}} |j\rangle_{E'}, \quad \langle \vartheta|_{j,A'} = \sqrt{(-1)^{n_j}} \langle j|_{A'}, \quad (116)$$

and $\vec{n} = \{n_1, n_2, n_3\}$, $n_1 = 0, n_2, n_3 \in \{0,1,2,3\}$.

Using these rank-one Kraus matrices $\sum_i N_i^{(A')}$, channel $\mathcal{N}_{AE'}$ can be rewritten as

$$\mathcal{N}_{AE'}(\rho_{A'}) = \mathcal{N}_{AE} \circ \mathcal{D}^{E \to E'} = \sum_i N_i^{(A')} \rho_{A'} N_i^{(A')\dagger}, \quad (117)$$

where $\sum_i N_i^{(A')\dagger} N_i^{(A')} = I$, from which follows that the degraded complementary channel $\mathcal{N}_{AE'}$, has become an entanglement-breaking channel – in other words, the $\mathcal{D}^{E \to E'}$ anti-degradable map given in (108) is, in fact, an entanglement-breaking map.

As follows, for a PD channel, neither the anti-degradable map $\mathcal{D}^{E \to E'}$ nor the complementary channel $\mathcal{N}_{AE'}$ can be expressed as the sum of rank-one Kraus operators.

∎

## 4 PD Channel Examples

### 4.1 Anti-degradable PD Channel

**Lemma 1** *There exists an anti-degradable qudit PD channel $\mathcal{N}$ that has entanglement-binding qudit complementary channel $\mathcal{N}_{AE}$.*

*Proof.*

Let the channel between Alice and Bob with respect to complementary channel $\mathcal{N}_{AE}$ be denoted by $\mathcal{N}_{AB}^{(AE)}$. Since $\mathcal{N}$ is anti-degradable, it follows that $\mathcal{N}_{AB}^{(AE)}$ is also anti-degradable.

First we show that the qudit channel $\mathcal{N}_{AB}^{(AE)}$ is anti-degradable. In the second part, we prove that the qudit complementary channel $\mathcal{N}_{AE}$ is degradable. Finally show that the degradable qudit complementary channel $\mathcal{N}_{AE}$ is entanglement-binding.

*Part 1*

Channel $\mathcal{N}$ has to be a qudit channel; this requirement follows from the fact that bound entanglement could not exist in a system that has total dimension less than or equal to 6 [10]. Using our notations, it means that $d_A d_E > 6$ has to hold for the dimension of input system $d_A$ and the dimension of the environment state $d_E$, in other words, $\mathcal{N}_{AB}^{(AE)}$ and $\mathcal{N}_{AE}$ have to be qudit (dim>2) channels.

The qudit channel $\mathcal{N}$ is defined with the following input and output spaces:
$$d_A = 4, \tag{118}$$
$$d_B = 3d_A = 12 \tag{119}$$
and
$$d_E = 2d_A = 8. \tag{120}$$
As follows, for channel $\mathcal{N}$, relation $d_E < d_B$ holds; i.e., it is a channel with small environment [10]. The CPTP maps of channel $\mathcal{N}$ are defined as follows:
$$\mathcal{N}_{AB}^{(AE)} : M_{d_A} \mapsto M_{d_B} = M_4 \mapsto M_{12}, \tag{121}$$
$$\mathcal{N}_{AE} : M_{d_A} \mapsto M_{d_E} = M_4 \mapsto M_8, \tag{122}$$
where $M_d$ are $d$ dimensional matrices. The CPTP map of the anti-degradable channel $\mathcal{N}_{AB}^{(AE)}$ assuming input density matrix $\rho$ can be rewritten as follows:
$$\begin{aligned}\mathcal{N}_{AB}^{(AE)}(\rho) &= x|0\rangle\langle 0| \otimes Tr(\rho) + (1-x)|1\rangle\langle 1| \otimes \mathcal{M}_{AB}^{(AE)}(\rho) \\ &= xTr(\rho) \oplus (1-x)\mathcal{M}_{AB}^{(AE)}(\rho),\end{aligned} \tag{123}$$
where $\frac{1}{2} \leq x \leq 1$, and $\mathcal{M}_{AB}^{(AE)}(\rho)$ is a CPTP map defined as
$$\mathcal{M}_{AB}^{(AE)} : M_4 \mapsto M_6, \tag{124}$$
hence channel $\mathcal{N}_{AB}^{(AE)}$ can be rewritten as follows:
$$\mathcal{N}_{AB}^{(AE)} : M_{d_A} \mapsto M_{d_B} = M_4 \mapsto M_2 \otimes M_6 = M_{12}. \tag{125}$$

Since for any $\frac{1}{2} \leq x \leq 1$ and CPTP map $\mathcal{M}_{AB}^{(AE)} : M_4 \mapsto M_6$, the channel $\mathcal{N}_{AB}^{(AE)}$ cannot be degradable; therefore, it trivially follows that the qudit channel $\mathcal{N}$ is anti-degradable [10].

Since channel $\mathcal{N}_{AB}^{(AE)}$ is anti-degradable, there is no existing CPTP degrading map $\mathcal{D}^{B \to E} : M_{d_B} \mapsto M_{d_E} = M_{12} \mapsto M_8$ for the channel, and $\mathcal{N}_{AB}^{(AE)} \circ \mathcal{D}^{B \to E}$ is also anti-degradable. To see it, we define the isometries

$$U : A \mapsto BE \tag{126}$$

and

$$L : B \mapsto CD, \tag{127}$$

where $B$ and $C$ are the outputs, while $E$ and $D$ are the environments for $\mathcal{N}_{AB}^{(AE)}$ and $\mathcal{D}^{B \to E}$, respectively.

For the maps $\mathcal{N}_{AB}^{(AE)}$, $\left(\mathcal{N}_{AB}^{(AE)}\right)^{-1}$, $\mathcal{D}^{B \to E}$ and $\left(\mathcal{N}_{AB}^{(AE)} \circ \mathcal{D}^{B \to E}\right)^{\perp}$, where $\perp$ stands for the complement, the following relation holds [10]:

$$\mathcal{N}_{AB}^{(AE)} : \mathcal{B}(\mathcal{H}_A) \mapsto \mathcal{B}(\mathcal{H}_B) \tag{128}$$

$$\left(\mathcal{N}_{AB}^{(AE)}\right)^{-1} : E \mapsto B, \tag{129}$$

$$\left(\mathcal{N}_{AB}^{(AE)}\right)^{\perp} \circ \left(\mathcal{D}^{B \to E}\right)^{-1} = \mathcal{N}_{AB}^{(AE)} : \mathcal{B}(\mathcal{H}_A) \mapsto \mathcal{B}(\mathcal{H}_B) \tag{130}$$

$$\mathcal{D}^{B \to E} : \mathcal{B}(\mathcal{H}_B) \mapsto \mathcal{B}(\mathcal{H}_C), \tag{131}$$

$$\left(\mathcal{N}_{AB}^{(AE)} \circ \mathcal{D}^{B \to E}\right)^{\perp} : \mathcal{B}(\mathcal{H}_A) \mapsto \mathcal{B}(\mathcal{H}_{DE}), \tag{132}$$

and

$$\left(\mathcal{N}_{AB}^{(AE)} \circ \mathcal{D}^{B \to E}\right)^{\perp}(\rho) = Tr_C \left(L \otimes I_E\right) U \rho U^{\dagger} \left(L^{\dagger} \otimes I_E\right), \tag{133}$$

where $\mathcal{B}(\cdot)$ are the set of bounded linear operators [10, 13].

Tracing out system $D$ from the complement of channel of degraded channel $\mathcal{N}_{AB}^{(AE)} \circ \mathcal{D}^{B \to E}$ leads to

$$\begin{aligned} Tr_D \left(\mathcal{N}_{AB}^{(AE)} \circ \mathcal{D}^{B \to E}\right)^{\perp}(\rho) &= Tr_{CD} U \rho U^{\dagger} \left(L \otimes I_E\right)\left(L^{\dagger} \otimes I_E\right) \\ &= Tr_C U \rho U^{\dagger} \\ &= \left(\mathcal{N}_{AB}^{(AE)}\right)^{\perp}(\rho), \end{aligned} \tag{134}$$

from which

$$\left(\mathcal{N}_{AB}^{(AE)} \circ \left(\mathcal{D}^{B\to E}\right)^{-1} \circ Tr_D\right)\left(\mathcal{N}_{AB}^{(AE)} \circ \mathcal{D}^{B\to E}\right)^{\perp}(\rho) = \mathcal{N}_{AB}^{(AE)} \circ \mathcal{D}^{B\to E}(\rho), \quad (135)$$

follows [10], which demonstrates that the PD channel $\mathcal{N}_{AB}^{(AE)} \circ \mathcal{D}^{B\to E}$ is, in fact, anti-degradable.

These results conclude Part 1 of the proof of Lemma 1.

□

*Part 2*

In the second part of this proof, we show that the complementary channel $\mathcal{N}_{AE} : M_{d_A} \mapsto M_{d_E} = M_4 \mapsto M_8$ is degradable.

The complementary channel $\mathcal{N}_{AE}$ is defined as follows:

$$\mathcal{N}_{AE}(\rho) = x|0\rangle\langle 0| \otimes I_4(\rho) + (1-x)|1\rangle\langle 1| \otimes \mathcal{M}_{AE}(\rho), \quad (136)$$

where $I_4$ is the four-dimensional identity matrix, $\frac{1}{2} \leq x \leq 1$, and $\mathcal{M}_{AE}(\rho)$ is a four-dimensional *entanglement-binding* [14] CPTP map

$$\mathcal{M}_{AE} : M_d \mapsto M_d = M_4 \mapsto M_4, \quad (137)$$

which is defined by the $N_i^{AE}$ Kraus operators as follows:

$$\begin{aligned}
N_0^{AE} &= \sqrt{\tfrac{1}{\sqrt{2}+2}} I \otimes |0\rangle\langle 0|, \\
N_1^{AE} &= \sqrt{\tfrac{1}{\sqrt{2}+2}} Z \otimes |1\rangle\langle 1|, \\
N_3^{AE} &= \sqrt{\tfrac{1}{2(\sqrt{2}+2)}} Z \otimes Y, \\
N_4^{AE} &= \sqrt{\tfrac{1}{2(\sqrt{2}+2)}} I \otimes X, \\
N_5^{AE} &= \sqrt{1-\tfrac{1}{\sqrt{2}+1}} X \otimes \begin{pmatrix} \tfrac{\sqrt{\sqrt{2}+2}}{2} & 0 \\ 0 & \tfrac{\sqrt{-\sqrt{2}+2}}{2} \end{pmatrix}, \\
N_6^{AE} &= \sqrt{1-\tfrac{1}{\sqrt{2}+1}} Y \otimes \begin{pmatrix} \tfrac{\sqrt{-\sqrt{2}+2}}{2} & 0 \\ 0 & \tfrac{\sqrt{\sqrt{2}+2}}{2} \end{pmatrix},
\end{aligned} \quad (138)$$

where X, Y and Z are the Pauli matrices.

As follows, the complementary channel $\mathcal{N}_{AE}(\rho)$ can be rewritten as follows:

$$\mathcal{N}_{AE}(\rho) : M_{d_A} \mapsto M_{d_E} = M_4 \mapsto M_2 \otimes M_4 = M_8. \quad (139)$$

and it is degradable, since for $\frac{1}{2} \leq x \leq 1$ and $\mathcal{M}_{AE} : M_4 \mapsto M_4$ is a CPTP map.

From the complementary channel $\mathcal{N}_{AE}$, the channel $\mathcal{N}_{AB}^{(AE)}$ can be simulated by the degradation map $\mathcal{D}^{E\to B}$. The map $\mathcal{D}^{E\to B}$ is a CPTP map, defined as

$$\mathcal{D}^{E\to B}: M_{d_E} \mapsto M_{d_B} = M_8 \mapsto M_{12}, \tag{140}$$

and the action of $\mathcal{D}^{E\to B}$ on product states $\varsigma_1 \otimes \varsigma_2$ can be expressed as

$$\begin{aligned}\mathcal{D}^{E\to B}\left(\varsigma_1 \otimes \varsigma_2\right) &= \tfrac{1-x}{x}\langle 0,\varsigma_1 0\rangle \mathcal{M}_{AB}^{(AE)}\left(\varsigma_1\right) \oplus \left(\tfrac{2x-1}{x}\langle 0,\varsigma_1 0\rangle Tr\left(\varsigma_2\right) + \langle 1,\varsigma_1 1\rangle Tr\left(\varsigma_2\right)\right) \\ &= \tfrac{1-x}{x}\mathcal{M}_{AB}^{(AE)}\left(\varsigma_1\right) \oplus \left(\tfrac{2x-1}{x}Tr\left(\varsigma_1\right) + Tr\left(\varsigma_2\right)\right).\end{aligned} \tag{141}$$

The set of $N_{j,k}^{EB}$, $j=\{0...5\}$, $k=\{0,1,2\}$ of Kraus operators of the degrading map $\mathcal{D}^{E\to B}$ is as follows:

$$\begin{aligned}&\left\{|0\rangle\langle 1||j\rangle, \sqrt{\tfrac{1-x}{x}}|1\rangle\langle 0|\otimes N_i^{AB}, \sqrt{\tfrac{2x-1}{x}}|0\rangle\langle 0||j\rangle\right\}_{j=0}^{d-1} = \\ &\left\{|0\rangle\langle 1||0\rangle, \sqrt{\tfrac{1-x}{x}}|1\rangle\langle 0|\otimes N_i^{AB}, \sqrt{\tfrac{2x-1}{x}}|0\rangle\langle 0||0\rangle\right\}, \\ &\left\{|0\rangle\langle 1||1\rangle, \sqrt{\tfrac{1-x}{x}}|1\rangle\langle 0|\otimes N_i^{AB}, \sqrt{\tfrac{2x-1}{x}}|0\rangle\langle 0||1\rangle\right\}, \\ &\qquad\qquad\vdots \\ &\left\{|0\rangle\langle 1||5\rangle, \sqrt{\tfrac{1-x}{x}}|1\rangle\langle 0|\otimes N_i^{AB}, \sqrt{\tfrac{2x-1}{x}}|0\rangle\langle 0||5\rangle\right\},\end{aligned} \tag{142}$$

where $\tfrac{1}{2} \leq x \leq 1$, and $N_i^{AB}$ are the Kraus operators of channel $\mathcal{M}_{AB}^{(AE)}$. One can readily check that, for product state input $\tau_1 \otimes \tau_2$

$$\begin{aligned}&\mathcal{N}_{AE}\left(\tau_1 \otimes \tau_2\right) \circ \mathcal{D}^{E\to B}\left(\varsigma_1 \otimes \varsigma_2\right) \\ &= \left(\mathcal{N}_{AE}\left(\tau_1 \otimes \tau_2\right)\right) \circ \left(\tfrac{1-x}{x}\mathcal{M}_{AB}^{(AE)}\left(\varsigma_1\right) \oplus \left(\tfrac{2x-1}{x}Tr\left(\varsigma_1\right) + Tr\left(\varsigma_2\right)\right)\right) \\ &= \mathcal{N}_{AB}^{(AE)}\left(\tau_1 \otimes \tau_2\right),\end{aligned} \tag{143}$$

where $\varsigma_1 \otimes \varsigma_2 = \mathcal{N}_{AE}\left(\tau_1 \otimes \tau_2\right)$, which proves that the complementary channel $\mathcal{N}_{AE}$ is degradable [10].

These results conclude Part 2 of the proof of Lemma 1.

$\square$

*Part 3*

In the third part of the proof, we show that the degradable complementary channel $\mathcal{N}_{AE}: M_{d_A} \mapsto M_{d_E} = M_4 \mapsto M_8$ is entanglement-binding.

The required preliminary condition on the entanglement-binding property is already satisfied, since $d_A d_E = 4 \cdot 6 > 6$; i.e., $\mathcal{N}$ is a qudit channel. To demonstrate the entanglement-binding property of the complementary channel $\mathcal{N}_{AE}$, we prove that map $\mathcal{M}_{AE} : M_4 \mapsto M_4$ is entanglement-binding. Let the input $|\Psi_{AB}\rangle$ of the channel $\mathcal{M}_{AE}$ represent half $A'$ of the $d_{AA'} = 4$, which is denoted by $d_A = 4$ dimensional maximally entangled system

$$|\Psi_{AA'}\rangle = \tfrac{1}{\sqrt{4}} \sum_{i=0}^{d_A-1} |i\rangle|i\rangle, \tag{144}$$

and the output of the channel $\mathcal{M}_{AE}$ is a four-dimensional bound-entangled system $\sigma$ as follows:

$$\sigma = (I \otimes \mathcal{M}_{AE})|\Psi_{AA'}\rangle, \tag{145}$$

acting over $\mathbb{C}^{d_A \cdot d_E} = \mathbb{C}^{4 \cdot 4} = \mathbb{C}^{16}$, which can be rewritten as

$$\sigma = p |\Psi_{AA'}\rangle\langle\Psi_{AA'}|_r + (1-p)\tfrac{1}{r^2-1}(I - |\Psi_{AA'}\rangle\langle\Psi_{AA'}|), \tag{146}$$

where $|\Psi_{AA'}\rangle\langle\Psi_{AA'}|_r$ is the maximally entangled state with a Schmidt rank of $r$, and $p \leq \tfrac{1}{r}$.

For channel $\mathcal{M}_{AE}$, the following two conditions have to hold: $\mathcal{M}_{AE}$ has to be a channel with $Q(\mathcal{M}_{AE}) = 0$, and channel output $\sigma$ has to be a *bound-entangled* state. The output system $\sigma = (I \otimes \mathcal{M}_{AE})|\Psi_{AA'}\rangle$ cannot constitute distillable entanglement, since $Q(\mathcal{M}_{AE}) = 0$ by assumption. On the other hand, from a separable output $\sigma = (I \otimes \mathcal{M}_{AE})|\Psi_{AA'}\rangle$, one also cannot produce the bound-entangled output system $\sigma$ by any LOCC action. The output $\sigma = (I \otimes \mathcal{M}_{AE})|\Psi_{AA'}\rangle$ for entangled input $|\Psi_{AA'}\rangle$ also cannot be a separable system, since a bound-entangled system cannot be produced from a separable state by LOCC [14,15].

These results conclude Part 3 of the proof of Lemma 1.

□
■

**Corollary 5** (On the degradation of a complementary channel). *For any anti-degradable PD channel there exists an anti-degradable but non entanglement-breaking map $\mathcal{D}^{E \to E'}$, that results in $\mathcal{N}_{AE'} = \mathcal{N}_{AE} \circ \mathcal{D}^{E \to E'}$. The resulting channel $\mathcal{N}_{AE'}$ is anti-degradable.*

*Proof.*
First we prove that $\mathcal{D}^{E \to E'}$ exists, then we show that resulting channel $\mathcal{N}_{AE'} = \mathcal{N}_{AE} \circ \mathcal{D}^{E \to E'}$ is anti-degradable.
The anti-degradable but non entanglement-breaking CPTP map $\mathcal{D}^{E \to E'}$ is defined as follows:

$$\mathcal{D}^{E\to E'}: M_{d_E} \mapsto M_{d_{E'}} = M_8 \mapsto M_2. \tag{147}$$

The map $\mathcal{D}^{E\to E'}(\rho)$ is given by Kraus operators as

$$\mathcal{D}^{E\to E'}(\rho) = \sum_i N_i \rho N_i^\dagger, \tag{148}$$

with $\sum_i N_i^\dagger N_i = I_8$, where $I_8$ is the eight-dimensional identity matrix, and the $2\times 8$-size Kraus matrices are

$$N_0 = \begin{pmatrix} \sqrt{a_1} & 0 & 0 & \ldots & 0 \\ 0 & \sqrt{a_2} & 0 & \ldots & 0 \end{pmatrix}, \tag{149}$$

$$N_1 = \begin{pmatrix} \sqrt{1-a_1} & 0 & 0 & \ldots & 0 \\ 0 & \sqrt{1-a_2} & 0 & \ldots & 0 \end{pmatrix}, \tag{150}$$

where $\sum_i a_i^2 = 1$, and

$$N_1^\dagger N_1 = I_8 - N_0^\dagger N_0 = \begin{pmatrix} 1-a_1 & 0 & 0 & \ldots & 0 \\ 0 & 1-a_2 & 0 & \ldots & 0 \\ 0 & 0 & 1 & \ldots & 0 \\ \vdots & \vdots & \vdots & \ddots & 0 \\ 0 & 0 & 0 & 0 & 1 \end{pmatrix}. \tag{151}$$

Since the input of the degrading map $\mathcal{D}^{E\to E'}$ has dimension $d_E = 8$, it also follows that the output $E'$ of the map $\mathcal{D}^{E\to E'}$ cannot be degradable. Matrix $N_1$ has size $2 \cdot d_E = 16$, hence the rank of $N_1 N_1^\dagger$ cannot be at most 2, which would be a required condition for degradability [10]. In (151), $N_1^\dagger N_1$ could have a rank of, at most, 2 only if $d_E \leq 3$, which is not satisfied, since $d_E = 8$. In other words, channel $\mathcal{N}_{AE'} = \mathcal{N}_{AE} \circ \mathcal{D}^{E\to E'}$ could be degradable if only if the dimension of $E$ of $\mathcal{D}^{E\to E'}$ would be $d_E \leq 3$, and its Choi rank $d_H$ (see Fig. 8 for the description of the output system) would be at most 2 [9,10,13]. From these results it can be concluded that channel $\mathcal{N}_{AE'} = \mathcal{N}_{AE} \circ \mathcal{D}^{E\to E'}$ is anti-degradable. The dimension of the degraded environment state $E'$ is $d_G = d_{E'} = 2$.

∎

**Lemma 2** (On the partial simulation property). *For any anti-degradable PD channel $\mathcal{N}$ there exists a degrading map $\mathcal{D}^{B\to E'}$ that can be used to simulate degraded environment state $E'$ from channel output B.*

*Proof.*

Let $\mathcal{N}_{AB}^{(AE')}$ denote the channel between Alice and Bob with respect to the degraded complementary channel $\mathcal{N}_{AE'} = \mathcal{N}_{AE} \circ \mathcal{D}^{E \to E'}$. We show that $\mathcal{N}_{AB}^{(AE')}$ is degradable, since it has the anti-degradable complementary channel $\mathcal{N}_{AE'} = \mathcal{N}_{AE} \circ \mathcal{D}^{E \to E'}$.

Since $\mathcal{N}_{AE'}$ is anti-degradable, using (136) it can be defined as follows:

$$\mathcal{N}_{AE'}(\rho) \mapsto x|0\rangle\langle 0| + (1-x)|1\rangle\langle 1|, \tag{152}$$

where $0 \leq x < \frac{1}{2}$, and $\mathcal{N}_{AE'}$ is as follows

$$\mathcal{N}_{AE'} : M_{d_A} \mapsto M_{d_{E'}} = M_4 \mapsto M_2, \tag{153}$$

and channel $\mathcal{N}_{AB}^{(AE')}$ using (123) can be expressed as

$$\begin{aligned}\mathcal{N}_{AB}^{(AE')}(\rho) &= x|0\rangle\langle 0| \otimes Tr(\rho) + (1-x)|1\rangle\langle 1| \otimes \mathcal{M}_{AB}^{(AE)}(\rho) \\ &= x \cdot Tr(\rho) \oplus (1-x) \cdot \mathcal{M}_{AB}^{(AE)}(\rho),\end{aligned} \tag{154}$$

where $0 \leq x < \frac{1}{2}$. Channel $\mathcal{N}_{AB}^{(AE')}$ is defined as

$$\mathcal{N}_{AB}^{(AE')} : M_{d_A} \mapsto M_{d_B} = M_4 \mapsto M_{12}, \tag{155}$$

i.e., $\mathcal{N}_{AB}^{(AE')}$ is just the same channel as $\mathcal{N}_{AB}^{(AE)}$; however, its complementary channel is $\mathcal{N}_{AE'} = \mathcal{N}_{AE} \circ \mathcal{D}^{E \to E'}$ instead of $\mathcal{N}_{AE}$, and in (154) it is defined by $0 \leq x < \frac{1}{2}$, instead of $\frac{1}{2} \leq x \leq 1$, as in (123). The complementary channel $\mathcal{N}_{AE'}$ is anti-degradable; and, channel $\mathcal{N}_{AB}^{(AE')}$ has become available to simulate $\mathcal{N}_{AE'}$ by $\mathcal{D}^{B \to E'}$ if $\mathcal{D}^{B \to E'} = \left(\mathcal{N}_{AB}^{AE'}\right)^{-1} \circ \mathcal{N}_{AE} \circ \mathcal{D}^{E \to E'}$ holds; see (95). An important conclusion immediately follows: from now on, the channel output system $B$ can be used to simulate the degraded environment state $E'$ outputted by $\mathcal{N}_{AE'} = \mathcal{N}_{AE} \circ \mathcal{D}^{E \to E'}$.

∎

**Lemma 3** (The PD property). *For any PD channel $\mathcal{N}$, there exists $\mathcal{D}^{B \to E'}$ for $\mathcal{N}_{AB}^{(AE')}$ that results in $\mathcal{N}_{AB'}^{(AE')} = \mathcal{N}_{AB}^{(AE')} \circ \mathcal{D}^{B \to E'} = \mathcal{N}_{AE'} = \mathcal{N}_{AE} \circ \mathcal{D}^{E \to E'}$.*

*Proof.*

Let $\mathcal{N}_{AB}^{(AE')}$ denote the channel between Alice and Bob with respect to the degraded complementary channel $\mathcal{N}_{AE'} = \mathcal{N}_{AE} \circ \mathcal{D}^{E \to E'}$. Let $\mathcal{N}_{AB'}^{(AE')}$ refer to the degraded channel $\mathcal{N}_{AB'}^{(AE')} = \mathcal{N}_{AB}^{(AE')} \circ \mathcal{D}^{B \to E'}$. In the proof of Lemma 2, we have already seen that $\mathcal{N}_{AE'} = \mathcal{N}_{AE} \circ \mathcal{D}^{E \to E'}$ is anti-degradable. We show that there exists a degrading map, $\mathcal{D}^{B \to E'}$, for channel $\mathcal{N}_{AB}^{(AE')}$ that can be used to simulate the degraded environment $E'$ from channel output $B$.

The degradation map $\mathcal{D}^{B \to E'}$ is a CPTP map
$$\mathcal{D}^{B \to E'} : M_{d_B} \mapsto M_{d_{E'}} = M_{12} \mapsto M_2, \tag{156}$$

whose action on product input states $\varsigma_1 \otimes \varsigma_2$ is

$$\begin{aligned}\mathcal{D}^{B \to E'}(\varsigma_1 \otimes \varsigma_2) &= \tfrac{1-x}{x}\langle 0, \varsigma_1 0\rangle \mathcal{N}_{AE'}(\varsigma_1) \oplus \left(\tfrac{2x-1}{x}\langle 0, \varsigma_1 0\rangle Tr(\varsigma_2) + \langle 1, \varsigma_1 1\rangle Tr(\varsigma_2)\right) \\ &= \tfrac{1-x}{x}\mathcal{N}_{AE'}(\varsigma_1) \oplus \left(\tfrac{2x-1}{x} Tr(\varsigma_1) + Tr(\varsigma_2)\right),\end{aligned} \tag{157}$$

where $\mathcal{N}_{AE'}(\rho) \mapsto x|0\rangle\langle 0| + (1-x)|1\rangle\langle 1|$ is the degraded complementary channel $\mathcal{N}_{AE'} : M_{d_A} \mapsto M_{d_{E'}} = M_4 \mapsto M_2$ as defined in (152), with $0 \leq x < \tfrac{1}{2}$.

The set of Kraus operators $N_{j,k}^{BE'}$, $j = \{0 \ldots d_{E'} - 1\}$, $k = \{0, 1, 2\}$ of degrading map $\mathcal{D}^{B \to E'}$ are defined as follows:

$$\begin{aligned}&\left\{|0\rangle\langle 1||j\rangle, \sqrt{\tfrac{1-x}{x}}|1\rangle\langle 0| \otimes N_i^{AE'}, \sqrt{\tfrac{2x-1}{x}}|0\rangle\langle 0||j\rangle\right\}_{j=0}^{d-1} = \\ &\left\{|0\rangle\langle 1||0\rangle, \sqrt{\tfrac{1-x}{x}}|1\rangle\langle 0| \otimes N_i^{AE'}, \sqrt{\tfrac{2x-1}{x}}|0\rangle\langle 0||0\rangle\right\}, \\ &\left\{|0\rangle\langle 1||1\rangle, \sqrt{\tfrac{1-x}{x}}|1\rangle\langle 0| \otimes N_i^{AE'}, \sqrt{\tfrac{2x-1}{x}}|0\rangle\langle 0||1\rangle\right\},\end{aligned} \tag{158}$$

where $\tfrac{1}{2} \leq x \leq 1$, and $N_i^{AE'}$ are the $2 \times 8$ Kraus matrices of the degraded complementary channel $\mathcal{N}_{AE'}$ as follows:

$$N_0^{AE'} = \begin{pmatrix} \sqrt{a_1} & 0 & 0 & \ldots & 0 \\ 0 & \sqrt{a_2} & 0 & \ldots & 0 \end{pmatrix}, \tag{159}$$

$$N_1^{AE'} = \begin{pmatrix} \sqrt{1-a_1} & 0 & 0 & \ldots & 0 \\ 0 & \sqrt{1-a_2} & 0 & \ldots & 0 \end{pmatrix}, \tag{160}$$

where $\sum_i a_i^2 = 1$ and $\sum_i N_i^\dagger N_i = I_8$.

One can readily check that, for product input states $\tau_1 \otimes \tau_2$

$$\begin{aligned}
\mathcal{N}_{AB'}^{(AE')} &= \mathcal{N}_{AB}^{(AE')}(\tau_1 \otimes \tau_2) \circ \mathcal{D}^{B \to E'}(\varsigma_1 \otimes \varsigma_2) \\
&= \left(\mathcal{N}_{AB}^{(AE')}(\tau_1 \otimes \tau_2)\right) \circ \left(\tfrac{1-x}{x}\mathcal{N}_{AE'}(\varsigma_1) \oplus \left(\tfrac{2x-1}{x}Tr(\varsigma_1) + Tr(\varsigma_2)\right)\right), \quad (161)\\
&= \mathcal{N}_{AE'}(\tau_1 \otimes \tau_2),
\end{aligned}$$

where $\varsigma_1 \otimes \varsigma_2 = \mathcal{N}_{AB}^{(AE')}(\tau_1 \otimes \tau_2)$.

The dimension $E'$ is $d_G = d_{E'} = 2$. From the definition of channel $\mathcal{N}$, we have $d_B = 12$, from which already follows that channel $\mathcal{N}_{AB'}^{(AE')} = \mathcal{N}_{AB}^{(AE')} \circ \mathcal{D}^{B \to E'}$ cannot be degradable independently from the Choi rank $d_F$ of $\mathcal{N}_{AB'}^{(AE')}$.

The $2 \times 12$-size Kraus matrices $N_0^{AB'}$ and $N_1^{AB'}$ of the degraded channel $\mathcal{N}_{AB'}^{(AE')}$ as follows:

$$N_0^{AB'} = \begin{pmatrix} \sqrt{a_1} & 0 & 0 & \dots & 0 \\ 0 & \sqrt{a_2} & 0 & \dots & 0 \end{pmatrix}, \quad (162)$$

$$N_1^{AB'} = \begin{pmatrix} \sqrt{1-a_1} & 0 & 0 & \dots & 0 \\ 0 & \sqrt{1-a_2} & 0 & \dots & 0 \end{pmatrix}, \quad (163)$$

where $\sum_i a_i^2 = 1$, $\sum_i N_i^\dagger N_i = I_{12}$, where $I_{12}$ is the twelve-dimensional identity matrix. The Kraus matrices $N_i^{AB'}$ of $\mathcal{N}_{AB'}^{(AE')}$ have size $2d_B = 24$, and $\left(N_1^{AB'}\right)^\dagger N_1^{AB'}$ is expressed as

$$\left(N_1^{AB'}\right)^\dagger N_1^{AB'} = I_{12} - \left(N_0^{AB'}\right)^\dagger N_0^{AB'} = \begin{pmatrix} 1-a_1 & 0 & 0 & \dots & 0 \\ 0 & 1-a_2 & 0 & \dots & 0 \\ 0 & 0 & 1 & \dots & 0 \\ \vdots & \vdots & \vdots & \ddots & 0 \\ 0 & 0 & 0 & 0 & 1 \end{pmatrix}, \quad (164)$$

for which the rank cannot be more than two. The $N_1 N_1^\dagger$ could have a rank of, at most, two if and only if $d_B \leq 3$, which is not satisfied, would hold. From these results, it can be concluded that channel $\mathcal{N}_{AB}^{(AE')}$ is degradable with respect to degraded complementary channel $\mathcal{N}_{AE'}$; and, by using $\mathcal{D}^{B \to E'}$, the following relationship holds:

$$\mathcal{N}_{AB'}^{(AE')} = \mathcal{N}_{AB}^{(AE')} \circ \mathcal{D}^{B \to E'} = \mathcal{N}_{AE'}, \quad (165)$$

which shows clearly that the anti-degradable channel $\mathcal{N}$ is, indeed, an anti-degradable PD channel.

∎

## 4.2 Symmetric PD Channel

**Lemma 4** *An anti-degradable $\mathcal{N}_{AB}^{AE'}$ with the degraded entanglement-binding complementary channel $\mathcal{N}_{AE'}$ formulate a symmetric channel.*

*Proof.*

Let us assume that $d_A = 4$ and $d_B = d_{E'} = 8$. For any PD channel, the degrading map $\mathcal{D}^{E \to E'}$ exists by assumption on the PD channel structure. The degrading map $\mathcal{D}^{E \to E'}$ makes $\mathcal{N}_{AE}$ an anti-degradable $\mathcal{N}_{AE'}$. If $\mathcal{N}_{AB}^{AE}$ is anti-degradable, the degraded channel $\mathcal{N}_{AE'}$ remains anti-degradable. The channel $\mathcal{N}_{AB}^{AE'}$ and $\mathcal{N}_{AE'}$ are defined as follows:

$$\mathcal{N}_{AB}^{AE'} : M_{d_A} \mapsto M_{d_B} = M_4 \mapsto M_8, \tag{166}$$

$$\mathcal{N}_{AE'} : M_{d_A} \mapsto M_{d_{E'}} = M_4 \mapsto M_8. \tag{167}$$

and

$$\mathcal{N}_{AB}^{AE'}(\rho) = Tr_{E'}\{V\rho V^{\dagger}\}, \tag{168}$$

$$\mathcal{N}_{AE'}(\rho) = Tr_B\{V\rho V^{\dagger}\}, \tag{169}$$

where

$$V : \mathbb{C}_8 \mapsto \mathbb{C}_{12} \tag{170}$$

is a partial isometry that maps onto the symmetric subspace of $BE$ with range of the symmetric subspace [10] of $\mathbb{C}_8 \mapsto \mathbb{C}_8$, from which follows that

$$\mathcal{N}_{AB}^{AE'}(\rho) = Tr_{E'}\{V\rho V^{\dagger}\} = \mathcal{N}_{AE'}(\rho) = Tr_B\{V\rho V^{\dagger}\}, \tag{171}$$

i.e., channel $\mathcal{N}$ *is symmetric PD channel.* For the degradation map of a symmetric PD channel, see (96).

∎

**Corollary 6** *The quantum capacity of any PD channel is additive in an arbitrary dimension.*

*Proof.*

From the results of Section 4, it follows that there exist both degradable and anti-degradable channel $\mathcal{N}$ that contains entanglement-binding complementary channel $\mathcal{N}_{AE}$ with an anti-degradable but non entanglement-breaking map, $\mathcal{D}^{E \to E'}$, from which the degraded complementary channel $\mathcal{N}_{AE'} = \mathcal{N}_{AE} \circ \mathcal{D}^{E \to E'}$ can be simulated. The complementary channel $\mathcal{N}_{AE'}$ is always anti-degradable. It is also proven that for an anti-degradable $\mathcal{N}_{AB}^{(AE)}$ with entanglement-binding $\mathcal{N}_{AE}$, there exists degraded complementary channel $\mathcal{N}_{AE'}$ such that the resulting $\mathcal{N}_{AB}^{(AE')}$ is a PD channel. For any PD channels it is possible to define degrading map $\mathcal{D}^{B \to E'}$, from which degraded environment $E'$ can be simulated by $B$, and that leads to $\mathcal{N}_{AB'}^{(AE')} = \mathcal{N}_{AB}^{(AE')} \circ \mathcal{D}^{B \to E'}$. For a symmetric PD channel the quantum capacity also additive, and the degradation maps $\mathcal{D}^{B \to E'} = \left(\mathcal{N}_{AB}^{AE'}\right)^{-1} \circ \mathcal{N}_{AE} \circ \mathcal{D}^{E \to E'}$ and $\eta^{E' \to B} = \left(\mathcal{N}_{AE} \circ \mathcal{D}^{E \to E'}\right)^{-1} \circ \mathcal{N}_{AB}^{AE'}$ both exist.

∎

# 5 Conclusions

We defined the structure of the partially degradable (PD) and conjugate-PD channel set. PD channels are similar fix points in the anti-degradable family than the Hadamard channels in the degradable set. PD channels could be found in the set of qudit channels that have entanglement-binding complementary channel. We proved that for an arbitrary dimensional PD quantum channel the quantum capacity is additive. The partial degradability property causes changes in the structure of the complementary channel that makes it possible to use the channel output for the simulation of the degraded environment. We called this property to partial simulation that is clearly not possible for an anti-degradable channel that has no the partial degradation map. We conjecture that the partial degradability property has many exciting still unrevealed benefits and possibilities for quantum communications.

# Acknowledgements

The results discussed above are supported by the grant TAMOP-4.2.2.*B*-10/1--2010-0009 and COST Action MP1006.

# Supplemental Information

In the Supplemental Information we analyze the performance of a PD channel for the transmission of quantum information.

## S.1 Performance of a PD channel

In this section we analyze the performance of a PD channel by using polar codes for the transmission of quantum information. We analyze the transmission rates of a degradable PD and anti-degradable PD channel. Finally, we compare the performance of a PD channel with non-PD degradable and anti-degradable channels. (*Note*: The results follow for the degradable/anti-degradable conjugate-PD channels.)

### S.1.1 Code construction

The $S_{in}^{degr}$ set of polar codewords that can transmit quantum information over the *degradable* (*but not PD*) channel $\mathcal{N}$ is defined as follows [S1]:

$$S_{in}^{degr}(\mathcal{N}) = \left(\mathcal{G}(\mathcal{N}_{amp},\beta) \cap \mathcal{G}(\mathcal{N}_{phase},\beta)\right)^{degr}, \tag{S.1}$$

where $\left|S_{in}^{degr}\right| = k$. The valuable quantum information will be transmitted by the set $S_{in}^{degr}(\mathcal{N})$ over the logical channels that can transmit both amplitude and phase information. All of the other input codewords cannot be used for quantum communication. These codewords will transmit only ancillary information (called *frozen* bits) for the decoding that cannot represent valuable quantum information between Alice and Bob. This set is defined by $S_{bad}(\mathcal{N})$, as follows:

$$\begin{aligned} S_{bad}(\mathcal{N}) = &\left(\mathcal{G}(\mathcal{N}_{amp},\beta) \cap \mathcal{B}(\mathcal{N}_{phase},\beta)\right) \\ &\cup \left(\mathcal{B}(\mathcal{N}_{amp},\beta) \cap \mathcal{G}(\mathcal{N}_{phase},\beta)\right) \\ &\cup \left(\mathcal{B}(\mathcal{N}_{amp},\beta) \cap \mathcal{B}(\mathcal{N}_{phase},\beta)\right), \end{aligned} \tag{S.2}$$

where $\left|S_{in}^{degr}\right| + \left|S_{bad}\right| = n = 2^k$. From set $S_{bad}$, we define the amplitude and phase frozen inputs as

$$\mathcal{B}(\mathcal{N}) = \mathcal{B}(\mathcal{N}_{amp},\beta) \cap \mathcal{B}(\mathcal{N}_{phase},\beta). \tag{S.3}$$

The density matrices of the frozen bits $\varsigma_{A_1},\ldots,\varsigma_{A_{n-l}}$ belong to the phase-frozen or amplitude-frozen sets

$$\mathcal{P}_1(\mathcal{N}) = \left(\mathcal{G}(\mathcal{N}_{amp},\beta) \cap \mathcal{B}(\mathcal{N}_{phase},\beta)\right)^{degr}, \tag{S.4}$$

$$\mathcal{P}_2(\mathcal{N}) = \left(\mathcal{B}(\mathcal{N}_{amp},\beta) \cap \mathcal{G}(\mathcal{N}_{phase},\beta)\right)^{degr}. \tag{S.5}$$

Sets $\mathcal{P}_1(\mathcal{N})$ and $\mathcal{P}_2(\mathcal{N})$ are disjoint [S1].

If channel $\mathcal{N}$ is assumed to be degradable but not PD, then the following $R_Q(\mathcal{N})$ rate can be achieved over its logical channel $\mathcal{N}_{AB}$:

$$\begin{aligned}R_Q(\mathcal{N}) &= \lim_{n\to\infty}\frac{1}{n}\left(\left|S_{in}^{degr}(\mathcal{N})\right|\right)\\ &= \lim_{n\to\infty}\frac{1}{n}\left|\mathcal{G}(\mathcal{N}_{amp},\beta) \cap \mathcal{G}(\mathcal{N}_{phase},\beta)\right|.\end{aligned} \tag{S.6}$$

*Note*: In the setting of polar coding, each capacity formula represents *symmetric* capacities: the input distribution is assumed to be *uniform*. In the case of $Q(\mathcal{N})$, which will be achieved by $R_Q(\mathcal{N})$, we also talk about the symmetric quantum capacity. Since the input distribution is assumed to be uniform, *no maximization is needed* in the general capacity formulas.

## On the selection and transmission of frozen bits

**Remark S.1** *In the encoding and decoding process, Alice and Bob have to use both the amplitude and the phase frozen states to recover the quantum information. In the encoding process Alice uses the $|0\rangle$ ancillary states as amplitude-frozen information, and the $\frac{1}{\sqrt{2}}(|0\rangle + |1\rangle)$ states as phase-frozen information. In the channel evolution process these ancillary states are transmitted via only $\mathcal{P}_1$, since the set $\mathcal{P}_2$ cannot be used for information transmission.*

*Note that Alice and Bob agreed that classical information can be transmitted only by set $\mathcal{G}(\mathcal{N}_{amp},\beta)$, and $\mathcal{P}_2$ does not belong to this set). In the decoding phase, Bob can easily synthesize the set $\mathcal{P}_2$ by applying a Hadamard operation $H^{\otimes|\Omega_{\mathcal{P}_1}|}(\Omega_{\mathcal{P}_1}) = \mathcal{P}_2$ on the corresponding subset $\Omega_{\mathcal{P}_1}$ of $\mathcal{P}_1$ to obtain the set $\mathcal{P}_2$ of amplitude-frozen bits for the decoding.*

**Remark S.2** *Since Alice and Bob use only the set of $\mathcal{G}(\mathcal{N}_{amp}, \beta)$ for classical communication, it follows that $\lim_{n \to \infty} \frac{1}{n} |\mathcal{P}_2(\mathcal{N})| = 0$ holds for both degradable and anti-degradable channels. For a degradable channel, the frozen bits will be transmitted via $\mathcal{P}_1(\mathcal{N})$, while for an anti-degradable channel the frozen bits are selected from $\mathcal{P}_1(\mathcal{N}) \cup \mathcal{B}(\mathcal{N})$. Since set $\mathcal{B}(\mathcal{N})$ requires pre-shared entanglement, for an anti-degradable channel the rate of entanglement consumption is non-zero. For a degradable channel, Alice and Bob do not have to use the set $\mathcal{B}(\mathcal{N})$, hence for a degradable channel no pre-shared entanglement required.*

For the frozen bits of any PD channel the following relations hold:
$$\lim_{n \to \infty} \frac{1}{n} \left| \left( \mathcal{P}_1 \setminus \mathcal{P}_1' \right) \right| > 0, \tag{S.7}$$

Since set $\mathcal{P}_2$ is not used for the transmission of frozen-bits, one obtains
$$\lim_{n \to \infty} \frac{1}{n} \left| \left( \mathcal{P}_2 \right) \right| = \lim_{n \to \infty} \frac{1}{n} \left| \left( \mathcal{P}_2 \setminus \mathcal{P}_2' \right) \right| = \lim_{n \to \infty} \frac{1}{n} \left| \left( \mathcal{P}_2' \right) \right| = 0, \tag{S.8}$$

that results in
$$\lim_{n \to \infty} \frac{1}{n} \left( \left| \left( \mathcal{P}_2 \setminus \mathcal{P}_2' \right) \right| + \left| \mathcal{P}_2' \right| \right) = 0. \tag{S.9}$$

In the decoding phase (S.8) is transformed into
$$\lim_{n \to \infty} \frac{1}{n} \left| H^{\otimes \left| \Omega_{\mathcal{P}_1 \setminus \mathcal{P}_1'} \right|} \left( \Omega_{\mathcal{P}_1 \setminus \mathcal{P}_1'} \right) \right| = \lim_{n \to \infty} \frac{1}{n} \left| \left( \mathcal{P}_2 \setminus \mathcal{P}_2' \right) \right| > 0, \tag{S.10}$$

where $H$ is the Hadamard operation.

### S.1.2 Degradable PD Channel

First, we assume that the quantum channel $\mathcal{N}$ is a degradable PD channel. For the quantum communication over $\mathcal{N}$, we use the good sets that are usable for channels that are degradable PD, but not for a standard degradable channel. The degradable conjugate-PD channels also belong to this set and the proof follows from [S2]. Sets $\mathcal{P}_1'(\mathcal{N})$ and $\mathcal{P}_2'(\mathcal{N})$ are in the sets of frozen bits defined for the degradable channel, as
$$\mathcal{P}_1'(\mathcal{N}) \subseteq \mathcal{P}_1(\mathcal{N}), \ \mathcal{P}_2'(\mathcal{N}) \subseteq \mathcal{P}_2(\mathcal{N}). \tag{S.11}$$

The amplitude- and phase-frozen density matrices $\varsigma_{A_1}, \ldots, \varsigma_{A_{n-l}}$ for a degradable PD channel will be selected from sets

$$\mathcal{P}_1(\mathcal{N}) \setminus \mathcal{P}_1'(\mathcal{N}) \tag{S.12}$$

and

$$\mathcal{P}_2(\mathcal{N}) \setminus \mathcal{P}_2'(\mathcal{N}), \tag{S.13}$$

where

$$\mathcal{P}_1(\mathcal{N}) = \left(\mathcal{G}(\mathcal{N}_{amp},\beta) \cap \mathcal{B}(\mathcal{N}_{phase},\beta)\right)^{degr}, \tag{S.14}$$

$$\mathcal{P}_2(\mathcal{N}) = \left(\mathcal{B}(\mathcal{N}_{amp},\beta) \cap \mathcal{G}(\mathcal{N}_{phase},\beta)\right)^{degr}, \tag{S.15}$$

with

$$\left|S_{bad}(\mathcal{N})\right| = n - (m + \Delta) = n - m - \left|\mathcal{P}_1'(\mathcal{M})\right|. \tag{S.16}$$

The improvement in the rate of quantum communication in comparison to the just degradable, but not PD version of the channel $\mathcal{N}$ is

$$\Delta = \left|\mathcal{P}_1(\mathcal{N})\right| - \left|\mathcal{P}_1(\mathcal{N}) \setminus \mathcal{P}_1'(\mathcal{N})\right| = \left|\mathcal{P}_1'(\mathcal{N})\right|, \tag{S.17}$$

which is equal to

$$\Delta = R_Q(\mathcal{N}_{AE}) - R_Q(\mathcal{N}_{AE'}) = \lim_{n \to \infty} \frac{1}{n}\left(\left|\mathcal{P}_1'\right|\right). \tag{S.18}$$

*Note*: One can readily check that for the conjugate degradable channels of [1], $\Delta = 0$ [S2].

**Theorem S.1** (On the property of the constructed polar code). *For a degradable PD channel, only sets $S_{in}^{PD,degr}$ and $\mathcal{P}_1(\mathcal{N}) \setminus \mathcal{P}_1'(\mathcal{N})$ have to be used. Set $S_{in}^{PD,degr}$ conveys the valuable quantum information, while set $\mathcal{P}_1(\mathcal{N}) \setminus \mathcal{P}_1'(\mathcal{N})$ transmits the phase frozen density matrices for the decoding.*

*Proof.*

For a degradable PD channel, an extended set of polar codes will be available for quantum communication, in comparison to the just degraded case. It will be referred to as

$$S_{in}^{PD,degr}(\mathcal{N}) = S_{in}^{degr}(\mathcal{N}) \cup \mathcal{P}_1'(\mathcal{N}) = \mathcal{G}(\mathcal{N}_{amp},\beta) \cap \mathcal{G}(\mathcal{N}_{phase},\beta), \tag{S.19}$$

with

$$\left|S_{in}^{PD,degr}\right| = \left|S_{in}^{degr}\right| + \Delta = m + \Delta, \tag{S.20}$$

where $m < m + \Delta$ [S2].

The polar code sets that can be used for capacity-achieving quantum communication over a degradable PD channel $\mathcal{N}$ are defined as follows. The set that can transmit valuable information in $\mathcal{N}$ is

$$S_{in}^{PD,degr}(\mathcal{N}) = \mathcal{G}(\mathcal{N}_{amp},\beta) \cap \mathcal{G}(\mathcal{N}_{phase},\beta), \tag{S.21}$$

while the frozen density matrices $\varsigma_{A_1},\ldots,\varsigma_{A_{n-l}}$ are selected form the sets $\mathcal{P}_1(\mathcal{N}) \setminus \mathcal{P}'_1(\mathcal{N})$ and $\mathcal{P}_2(\mathcal{N}) \setminus \mathcal{P}'_2(\mathcal{N})$, which sets can be rewritten in the following form:

$$\mathcal{P}_1(\mathcal{N}) \setminus \mathcal{P}'_1(\mathcal{N}) = \left(\mathcal{G}(\mathcal{N}_{amp},\beta) \cap \mathcal{B}(\mathcal{N}_{phase},\beta)\right)^{degr} \setminus \mathcal{P}'_1(\mathcal{N}) \tag{S.22}$$

and

$$\mathcal{P}_2(\mathcal{N}) \setminus \mathcal{P}'_2(\mathcal{N}) = \left(\mathcal{B}(\mathcal{N}_{amp},\beta) \cap \mathcal{G}(\mathcal{N}_{phase},\beta)\right)^{degr} \setminus \mathcal{P}'_2(\mathcal{N}), \tag{S.23}$$

hence sets $\mathcal{P}'_1(\mathcal{N})$ and $\mathcal{P}'_2(\mathcal{N})$ can be expressed as

$$\mathcal{P}'_1(\mathcal{N}) = \left(\mathcal{G}(\mathcal{N}_{amp},\beta) \cap \mathcal{B}(\mathcal{N}_{phase},\beta)\right)^{PD} \tag{S.24}$$

and

$$\mathcal{P}'_2(\mathcal{N}) = \left(\mathcal{B}(\mathcal{N}_{amp},\beta) \cap \mathcal{G}(\mathcal{N}_{phase},\beta)\right)^{PD}. \tag{S.25}$$

The remaining and unused set is defined as follows:

$$\mathcal{B}(\mathcal{N}) = \left(\mathcal{B}(\mathcal{N}_{amp},\beta) \cap \mathcal{B}(\mathcal{N}_{phase},\beta)\right)^{PD,degr}, \tag{S.26}$$

where

$$\begin{aligned}\mathcal{B}(\mathcal{N}) &= \left(\mathcal{B}(\mathcal{N}_{amp},\beta) \cap \mathcal{B}(\mathcal{N}_{phase},\beta)\right)^{PD,degr} \\ &= \left(\mathcal{B}(\mathcal{N}_{amp},\beta) \cap \mathcal{B}(\mathcal{N}_{phase},\beta)\right)^{degr} = \varnothing.\end{aligned} \tag{S.27}$$

Set $\mathcal{B}(\mathcal{N})$ is not used as frozen bits, since the channel $\mathcal{N}$ is degradable. Due to the difference between the codeword sets defined for a degradable channel and for a both degradable PD channel $\mathcal{N}$, the rate of quantum communication will also differ. If $|\mathcal{P}'_1| > 0$ condition holds, then the $R_Q$ rate of quantum communication for a degradable PD channel $\mathcal{N}$ with set $S_{in}^{PD,degr}$ is higher than the rate of quantum communication that can be obtained for a degradable channel with $S_{in}^{degr}$. For codewords sets $\mathcal{P}'_1$, $\mathcal{P}'_2$, $\mathcal{P}_1 \setminus \mathcal{P}'_1$ and $\mathcal{P}_2 \setminus \mathcal{P}'_2$ we have

$$\begin{aligned}&\lim_{n\to\infty} \frac{1}{n}\left|\mathcal{G}(\mathcal{N}_{amp},\beta)\right| + \\ &\lim_{n\to\infty} \frac{1}{n}\left|[n] \setminus \left((\mathcal{P}'_1 \cup \mathcal{P}'_2) \cup (\mathcal{P}_1 \setminus \mathcal{P}'_1) \cup (\mathcal{P}_2 \setminus \mathcal{P}'_2)\right)\right| = 1,\end{aligned} \tag{S.28}$$

with

$$\left|\mathcal{G}(\mathcal{N}_{amp},\beta)\setminus\mathcal{G}(\mathcal{N}_{amp},\beta)\cap\mathcal{G}(\mathcal{N}_{phs},\beta)\right|+\left|\mathcal{P}_2\setminus\mathcal{P}_2'\right|$$
$$+\left|[n]\setminus\left((\mathcal{P}_1'\cup\mathcal{P}_2')\cup(\mathcal{P}_1\setminus\mathcal{P}_1')\cup(\mathcal{P}_2\setminus\mathcal{P}_2')\right)\right|\leq n \quad (S.29)$$

and

$$([n]\setminus(\mathcal{P}_1'))\subseteq\left(\mathcal{G}(\mathcal{N}_{amp},\beta)\cap\mathcal{G}(\mathcal{N}_{phase},\beta)\right)^{PD}$$
$$\cup\left((\mathcal{P}_1\setminus\mathcal{P}_1')\right). \quad (S.30)$$

Assuming $\beta<0.5$, the following relation holds for the Holevo information of the logical channels $\mathcal{N}_{AB}$, $\mathcal{N}_{AE}$ and $\mathcal{N}_{AE'}$ of $\mathcal{N}$:

$$\chi(\mathcal{N}_{AB})=\lim_{n\to\infty}\frac{1}{n}\left(\left|\mathcal{G}(\mathcal{N}_{amp},\beta)\cup\mathcal{P}_2'\right|\right), \quad (S.31)$$

$$\chi(\mathcal{N}_{AE})=\lim_{n\to\infty}\frac{1}{n}\left(|\mathcal{P}_1|+|\mathcal{P}_2|\right), \quad (S.32)$$

$$\chi(\mathcal{N}_{AE'})=\lim_{n\to\infty}\frac{1}{n}\left(\left|(\mathcal{P}_1\setminus\mathcal{P}_1')\right|+\left|(\mathcal{P}_2\setminus\mathcal{P}_2')\right|\right). \quad (S.33)$$

For the degraded environment state $E'$ of $\mathcal{N}$, it results in

$$\chi(\mathcal{N}_{AE'})=\lim_{n\to\infty}\frac{1}{n}\left(\left|(\mathcal{P}_1\setminus\mathcal{P}_1')\right|\right), \quad (S.34)$$

and the rate of quantum communication over the channel $\mathcal{N}$ is (*Note:* no maximization and regularization needed in the first line of (S.35), since in terms of polar coding the quantum capacity is symmetric and it is additive for any PD channel.)

$$R_Q(\mathcal{N})=\chi(\mathcal{N}_{AB})-\chi(\mathcal{N}_{AE'})$$
$$=\lim_{n\to\infty}\frac{1}{n}\left(\left|\mathcal{G}(\mathcal{N}_{amp},\beta)\right|-\left|(\mathcal{P}_1\setminus\mathcal{P}_1')\right|\right). \quad (S.35)$$

The result obtained in (S.35) can be rewritten as follows:

$$R_Q(\mathcal{N})=\lim_{n\to\infty}\frac{1}{n}\left(\left|S_{in}^{degr}\right|+\left|\mathcal{P}_1'\right|\right)=\lim_{n\to\infty}\frac{1}{n}\left(\left|S_{in}^{PD,degr}\right|\right). \quad (S.36)$$

For a degradable PD channel we get:

$$R_Q(\mathcal{N})=\lim_{n\to\infty}\frac{1}{n}\left(\left|S_{in}^{degr}\right|+\left|\mathcal{P}_1'\right|\right)=$$
$$=\lim_{n\to\infty}\frac{1}{n}\left(\left|\mathcal{G}(\mathcal{N}_{amp},\beta)\right|-\left|(\mathcal{P}_1\setminus\mathcal{P}_1')\right|\right) \quad (S.37)$$
$$=\lim_{n\to\infty}\frac{1}{n}\left(\left|S_{in}^{PD,degr}\right|\right).$$

In other words, the available codewords for quantum communication over the degradable PD channel $\mathcal{N}$ is

$$\left|S_{in}^{PD,degr}\right| = \left|S_{in}^{degr}\right| + \left|\mathcal{P}_1'\right|. \tag{S.38}$$

These results show that for the non-empty set $\mathcal{P}_1'$, the results on rate of quantum communication $R_Q(\mathcal{N})$ for a degradable PD channel $\mathcal{N}$ exceed the result obtained for the just degradable version of the channel $\mathcal{N}$ with $\left|S_{in}^{degr}\right|$. In the decoding process, Bob synthesizes set $\left(\mathcal{P}_2 \setminus \mathcal{P}_2'\right)$ from $\left(\mathcal{P}_1 \setminus \mathcal{P}_1'\right)$ by applying a Hadamard operation on the corresponding subset $\Omega_{\mathcal{P}_1 \setminus \mathcal{P}_1'}$ of $\left(\mathcal{P}_1 \setminus \mathcal{P}_1'\right)$ to obtain the amplitude-frozen bits $H^{\otimes \left|\Omega_{\mathcal{P}_1 \setminus \mathcal{P}_1'}\right|} \left(\Omega_{\mathcal{P}_1 \setminus \mathcal{P}_1'}\right) = \left(\mathcal{P}_2 \setminus \mathcal{P}_2'\right)$.

The polar code constructed for quantum communication over a degradable PD channel $\mathcal{N}$ is summarized in Fig. S.1. The polar codeword set that can be used for quantum communication is depicted in grey. The further sets can be used only to transmit classical information in the form of frozen density matrices, or not used.

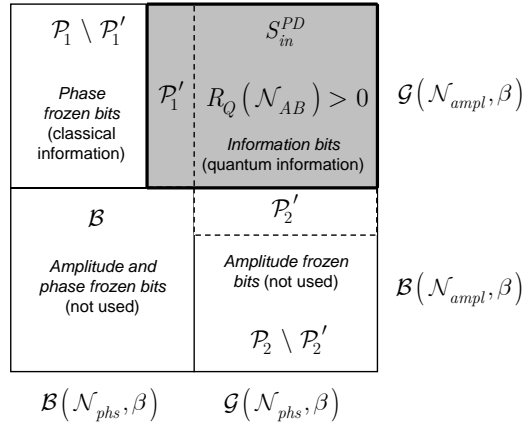

**Figure S.1**. The code structure of the polar code for quantum communication over a degradable PD channel. Set $S_{in}^{PD,degr}$ will be used for quantum communication. For the frozen bits only set $\left(\mathcal{P}_1 \setminus \mathcal{P}_1'\right)$ will be used, which set represents phase frozen density matrices. For the decoding, Bob synthesizes set $\left(\mathcal{P}_2 \setminus \mathcal{P}_2'\right)$ by applying a Hadamard operation on the corresponding subset $\Omega_{\mathcal{P}_1 \setminus \mathcal{P}_1'}$ of $\left(\mathcal{P}_1 \setminus \mathcal{P}_1'\right)$ to obtain the amplitude-frozen bits, $H^{\otimes \left|\Omega_{\mathcal{P}_1 \setminus \mathcal{P}_1'}\right|} \left(\Omega_{\mathcal{P}_1 \setminus \mathcal{P}_1'}\right) = \left(\mathcal{P}_2 \setminus \mathcal{P}_2'\right)$.

The code construction process for a degradable PD channel has resulted in six disjoint sets, $S_{in}^{PD,degr}(\mathcal{N})$, $\mathcal{P}_1(\mathcal{N})$, $\mathcal{P}_1(\mathcal{N}) \setminus \mathcal{P}_1'(\mathcal{N})$, $\mathcal{P}_2(\mathcal{N}) \setminus \mathcal{P}_2'(\mathcal{N})$, $\mathcal{P}_2'(\mathcal{N})$ and $\mathcal{B}(\mathcal{N})$, with relations

$$S_{in}^{PD,degr}\left(\mathcal{N}\right) \cap \left(\mathcal{P}_1\left(\mathcal{N}\right) \setminus \mathcal{P}_1'\left(\mathcal{N}\right)\right) \cap \left(\mathcal{P}_2\left(\mathcal{N}\right) \setminus \mathcal{P}_2'\left(\mathcal{N}\right)\right) \cap \mathcal{B}\left(\mathcal{N}\right) = \varnothing, \quad \text{(S.39)}$$

and

$$S_{in}^{PD,degr}\left(\mathcal{N}\right) \cup \left(\mathcal{P}_1\left(\mathcal{N}\right) \setminus \mathcal{P}_1'\left(\mathcal{N}\right)\right) \cup \left(\mathcal{P}_2\left(\mathcal{N}\right) \setminus \mathcal{P}_2'\left(\mathcal{N}\right)\right) \cup \mathcal{B}\left(\mathcal{N}\right) = [n]. \quad \text{(S.40)}$$

The sets $\mathcal{B}\left(\mathcal{N}\right)$, $\mathcal{P}_2\left(\mathcal{N}\right) \setminus \mathcal{P}_2'\left(\mathcal{N}\right)$ and $\mathcal{P}_2'\left(\mathcal{N}\right)$ will not be used in the transmission, hence these can be eliminated from the further calculations, i.e.,

$$S_{in}^{PD,degr}\left(\mathcal{N}\right) \cup \left(\mathcal{P}_1\left(\mathcal{N}\right) \setminus \mathcal{P}_1'\left(\mathcal{N}\right)\right) = [n]. \quad \text{(S.41)}$$

For the sets of valuable information and frozen bits, the relations

$$S_{in}^{PD,degr}\left(\mathcal{N}\right) \cap \left(\mathcal{P}_1\left(\mathcal{N}\right) \setminus \mathcal{P}_1'\left(\mathcal{N}\right)\right) \cap \left(\mathcal{P}_2\left(\mathcal{N}\right) \setminus \mathcal{P}_2'\left(\mathcal{N}\right)\right) = \varnothing \quad \text{(S.42)}$$

and

$$S_{in}^{PD,degr}\left(\mathcal{N}\right) \cup \left(\mathcal{P}_1\left(\mathcal{N}\right) \setminus \mathcal{P}_1'\left(\mathcal{N}\right)\right) = \left[2^k\right] \quad \text{(S.43)}$$

also follows from the code construction.

*These results conclude the proof of Theorem S.1.*

∎

### S.1.3 Anti-degradable PD Channel

For the quantum communication over an anti-degradable PD channel $\mathcal{N}$, we define the sets that are good for anti-degradable PD channels, but not for a standard anti-degradable channel. The anti-degradable conjugate-PD channels also belong to this set.

**Theorem S.2** (On the property of the constructed polar code). *For an anti-degradable PD channel, sets $S_{in}^{PD,anti\text{-}degr}$, $\mathcal{P}_1\left(\mathcal{N}\right) \setminus \mathcal{P}_1'\left(\mathcal{N}\right)$ and $\mathcal{B}\left(\mathcal{N}\right)$ have to be used in the transmission phase.*

*Proof.*
For an anti-degradable PD channel the improvement over non-PD anti-degradable channels is as follows:

$$\left|S_{in}^{PD,anti\text{-}degr}\right| = \left|S_{in}^{anti\text{-}degr}\right| + \left|\mathcal{P}_1'\right|, \quad \text{(S.44)}$$

$$\left|S_{in}^{PD,anti\text{-}degr}\right| = \left|S_{in}^{anti\text{-}degr}\right| + \Delta = m + \Delta, \quad \text{(S.45)}$$

where

$$S_{in}^{anti\text{-}degr}\left(\mathcal{N}\right) = \left(\mathcal{G}\left(\mathcal{N}_{amp},\beta\right) \cap \mathcal{G}\left(\mathcal{N}_{phase},\beta\right)\right) \setminus \mathcal{B}, \tag{S.46}$$

where $\left|S_{in}^{anti\text{-}degr}\right| = \left|S_{in}^{degr}\right| - \left|\mathcal{B}\right|$, and $m < m + \Delta$.

The boost in the rate of quantum communication in comparison to an anti-degradable non-PD channel $\mathcal{N}$ is

$$\Delta = \left|\mathcal{P}_1\left(\mathcal{N}\right)\right| - \left|\mathcal{P}_1\left(\mathcal{N}\right) \setminus \mathcal{P}_1'\left(\mathcal{N}\right)\right| = \left|\mathcal{P}_1'\left(\mathcal{N}\right)\right|, \tag{S.47}$$

that can be rewritten as

$$\Delta = R_Q\left(\mathcal{N}_{AE}\right) - R_Q\left(\mathcal{N}_{AE'}\right) = \lim_{n \to \infty} \frac{1}{n}\left|\mathcal{P}_1'\right|. \tag{S.48}$$

The frozen bits are transmitted via

$$\mathcal{P}_1\left(\mathcal{N}\right) \setminus \mathcal{P}_1'\left(\mathcal{N}\right), \tag{S.49}$$

and

$$\mathcal{B}\left(\mathcal{N}\right) \setminus \mathcal{B}'\left(\mathcal{N}\right), \tag{S.50}$$

where

$$\mathcal{P}_1'\left(\mathcal{N}\right) = \left(\mathcal{G}\left(\mathcal{N}_{amp},\beta\right) \cap \mathcal{B}\left(\mathcal{N}_{phase},\beta\right)\right)^{PD}, \tag{S.51}$$

$$\mathcal{P}_2'\left(\mathcal{N}\right) = \left(\mathcal{B}\left(\mathcal{N}_{amp},\beta\right) \cap \mathcal{G}\left(\mathcal{N}_{phase},\beta\right)\right)^{PD}, \tag{S.52}$$

$$\mathcal{B}\left(\mathcal{N}\right) = \left(\mathcal{B}\left(\mathcal{N}_{amp},\beta\right) \cap \mathcal{B}\left(\mathcal{N}_{phase},\beta\right)\right)^{PD,anti\text{-}degr}, \tag{S.53}$$

and

$$\begin{aligned}\mathcal{B}\left(\mathcal{N}\right) &= \left(\mathcal{B}\left(\mathcal{N}_{amp},\beta\right) \cap \mathcal{B}\left(\mathcal{N}_{phase},\beta\right)\right)^{PD,anti\text{-}degr} \\ &= \left(\mathcal{B}\left(\mathcal{N}_{amp},\beta\right) \cap \mathcal{B}\left(\mathcal{N}_{phase},\beta\right)\right)^{anti\text{-}degr} \neq \varnothing.\end{aligned} \tag{S.54}$$

The proof is similar to the previous case. If $\left|\mathcal{P}_1'\right| > 0$ condition holds, then the $R_Q$ rate of quantum communication for an anti-degradable PD channel $\mathcal{N}$ with set $S_{in}^{PD,anti\text{-}degr}$ is higher than the rate of quantum communication that can be obtained for an anti-degradable channel with $S_{in}^{anti\text{-}degr}$. For an anti-degradable PD channel the rate of entanglement consumption is greater then zero it follows that

$$\lim_{n \to \infty} \frac{1}{n}\left|\mathcal{B}\right| > 0, \tag{S.55}$$

where $\left|\left(\mathcal{P}_2 \setminus \mathcal{P}_2'\right) \cap \mathcal{G}\left(\mathcal{N}_{amp},\beta\right)\right| = 0$, $\left|\mathcal{P}_2' \cap \mathcal{G}\left(\mathcal{N}_{amp},\beta\right)\right| = 0$ and $\left|\mathcal{B}\left(\mathcal{N}_{amp},\beta\right) \cap \mathcal{G}\left(\mathcal{N}_{amp},\beta\right)\right| = 0$. In the code, the sets $\left(\mathcal{P}_1 \setminus \mathcal{P}_1'\right)$ and $\left(\mathcal{P}_2 \setminus \mathcal{P}_2'\right)$ belong to the

phase and amplitude frozen bits, $\left(\mathcal{P}_1 \setminus \mathcal{P}_1'\right) \subseteq \mathcal{B}\left(\mathcal{N}_{phase}, \beta\right)$, $\mathcal{P}_2 \subseteq \mathcal{B}\left(\mathcal{N}_{amp}, \beta\right)$, and $\mathcal{B}\left(\mathcal{N}_{amp}, \beta\right) = [n] \setminus \mathcal{G}\left(\mathcal{N}_{amp}, \beta\right)$. For the sets $\mathcal{P}_1'$ and $\mathcal{P}_2'$, the relations $\mathcal{P}_1' \cup \mathcal{P}_2' \subseteq \mathcal{G}\left(\mathcal{N}_{amp}, \beta\right) \cup \mathcal{G}\left(\mathcal{N}_{phase}, \beta\right)$, $\left(\mathcal{P}_1 \setminus \mathcal{P}_1'\right) \cup \mathcal{P}_2 \subseteq \mathcal{G}\left(\mathcal{N}_{amp}, \beta\right) \cup \mathcal{G}\left(\mathcal{N}_{phase}, \beta\right)$ hold. The sets $\mathcal{P}_1'$, $\mathcal{P}_2'$, $\mathcal{P}_1 \setminus \mathcal{P}_1'$, $\mathcal{P}_2 \setminus \mathcal{P}_2'$ and $\mathcal{B}$ are pairwise disjoint, and

$$
\begin{aligned}
&\lim_{n \to \infty} \frac{1}{n}\left|\mathcal{G}\left(\mathcal{N}_{amp}, \beta\right)\right| + \\
&\lim_{n \to \infty} \frac{1}{n}\left|\mathcal{P}_2\right| + \\
&\lim_{n \to \infty} \frac{1}{n}\left|\mathcal{B}\left(\mathcal{N}_{amp}, \beta\right) \cap \mathcal{B}\left(\mathcal{N}_{phs}, \beta\right)\right| \\
&= \lim_{n \to \infty} \frac{1}{n}\left(\left|S_{in}^{PD, anti\text{-}degr}\right| + \left|\left(\mathcal{P}_1 \setminus \mathcal{P}_1'\right)\right| + 2\left|\mathcal{B}\right|\right) = 1,
\end{aligned}
\tag{S.56}
$$

with

$$
\begin{aligned}
&\left|\mathcal{G}\left(\mathcal{N}_{amp}, \beta\right) \setminus \mathcal{G}\left(\mathcal{N}_{amp}, \beta\right) \cap \mathcal{G}\left(\mathcal{N}_{phs}, \beta\right)\right| \\
&+ \left|\left(\mathcal{P}_2 \setminus \mathcal{P}_2'\right)\right| \\
&+ \left|[n] \setminus \left(\left(\mathcal{P}_1' \cup \mathcal{P}_2'\right) \cup \left(\mathcal{P}_1 \setminus \mathcal{P}_1'\right) \cup \mathcal{P}_2\right)\right| \leq n.
\end{aligned}
\tag{S.57}
$$

Assuming $\beta < 0.5$, the following relation holds for the Holevo information of the logical channels $\mathcal{N}_{AB}$, $\mathcal{N}_{AE}$ and $\mathcal{N}_{AE'}$ of $\mathcal{N}$:

$$
\chi\left(\mathcal{N}_{AB}\right) = \lim_{n \to \infty} \frac{1}{n}\left(\left|\mathcal{G}\left(\mathcal{N}_{amp}, \beta\right) \cup \mathcal{P}_2'\right|\right), \tag{S.58}
$$

$$
\chi\left(\mathcal{N}_{AE}\right) = \lim_{n \to \infty} \frac{1}{n}\left(\left|\mathcal{P}_1\right| + \left|\mathcal{P}_2\right| + \left|\mathcal{B}\left(\mathcal{N}\right)\right|\right), \tag{S.59}
$$

$$
\chi\left(\mathcal{N}_{AE'}\right) = \lim_{n \to \infty} \frac{1}{n}\left(\left|\left(\mathcal{P}_1 \setminus \mathcal{P}_1'\right)\right| + \left|\left(\mathcal{P}_2 \setminus \mathcal{P}_2'\right)\right| + \left|\mathcal{B}\left(\mathcal{N}\right)\right|\right). \tag{S.60}
$$

For the degraded environment state $E'$ of $\mathcal{N}$, it results in

$$
\chi\left(\mathcal{N}_{AE'}\right) = \lim_{n \to \infty} \frac{1}{n}\left(\left|\left(\mathcal{P}_1 \setminus \mathcal{P}_1'\right)\right| + \left|\mathcal{B}\left(\mathcal{N}\right)\right|\right), \tag{S.61}
$$

and the rate of quantum communication over the anti-degradable PD channel $\mathcal{N}$ is

$$
\begin{aligned}
R_Q\left(\mathcal{N}\right) &= \chi\left(\mathcal{N}_{AB}\right) - \chi\left(\mathcal{N}_{AE'}\right) \\
&= \lim_{n \to \infty} \frac{1}{n}\left(\left|\mathcal{G}\left(\mathcal{N}_{amp}, \beta\right)\right| - \left|\left(\mathcal{P}_1 \setminus \mathcal{P}_1'\right)\right| - \left|\mathcal{B}\left(\mathcal{N}\right)\right|\right).
\end{aligned}
\tag{S.62}
$$

The result obtained in (S.35) can be rewritten as follows:

$$
R_Q\left(\mathcal{N}\right) = \lim_{n \to \infty} \frac{1}{n}\left(\left|S_{in}^{anti\text{-}degr}\right| + \left|\mathcal{P}_1'\right|\right) = \lim_{n \to \infty} \frac{1}{n}\left(\left|S_{in}^{PD, anti\text{-}degr}\right|\right). \tag{S.63}
$$

According to the polar coding scheme, for the $F$ fidelity parameter of frozen bit sets $\mathcal{P}_1 \setminus \mathcal{P}_1'$, $\mathcal{P}_2 \setminus \mathcal{P}_2'$ and $\mathcal{B}$ for the synthesized channels for $\beta < 0.5$:

$$\sqrt{F\left(S_{in}^{PD,\text{anti-degr}}\right)} < 2^{-n^\beta}, \tag{S.64}$$

and

$$\sqrt{F\left(\left(\mathcal{P}_1 \setminus \mathcal{P}_1'\right) \cup \left(\mathcal{P}_2 \setminus \mathcal{P}_2'\right) \cup \mathcal{B}\right)} \geq 1 - 2^{-n^\beta}. \tag{S.65}$$

If (S.64) and (S.65) hold, then $\left(\mathcal{P}_1 \setminus \mathcal{P}_1'\right) \cap \left(\mathcal{P}_2 \setminus \mathcal{P}_2'\right) \cap \mathcal{B} \neq \varnothing$. Sets $S_{in}^{\text{anti-degr}}$, $S_{in}^{PD,\text{anti-degr}}$, $\mathcal{P}_1' \subseteq \mathcal{P}_1$, $\mathcal{P}_2' \subseteq \mathcal{P}_2$ and $\mathcal{B}$ are disjoint with relations

$$\begin{aligned}
&\left|S_{in}^{\text{anti-degr}} \cup \mathcal{P}_1' \cup \mathcal{P}_2' \cup \left(\mathcal{P}_1 \setminus \mathcal{P}_1'\right) \cup \left(\mathcal{P}_2 \setminus \mathcal{P}_2'\right)\right| \\
&= \left|S_{in}^{\text{anti-degr}} \cup \mathcal{P}_1' \cup \left(\mathcal{P}_1 \setminus \mathcal{P}_1'\right) \cup \mathcal{P}_2\right| \\
&= \left|S_{in}^{PD,\text{anti-degr}} \cup \left(\mathcal{P}_1 \setminus \mathcal{P}_1'\right) \cup \left(\mathcal{P}_2 \setminus \mathcal{P}_2'\right)\right| \\
&= \left|S_{in}^{PD,\text{anti-degr}} \cup \left(\mathcal{P}_1 \setminus \mathcal{P}_1'\right)\right|,
\end{aligned} \tag{S.66}$$

hence, for an anti-degradable PD channel $\mathcal{N}$ we get:

$$\begin{aligned}
R_Q(\mathcal{N}) &= \lim_{n \to \infty} \frac{1}{n}\left(\left|S_{in}^{\text{anti-degr}}\right| + \left|\mathcal{P}_1'\right|\right) = \\
&= \lim_{n \to \infty} \frac{1}{n}\left(\left|\mathcal{G}\left(\mathcal{N}_{amp},\beta\right)\right| - \left|\left(\mathcal{P}_1 \setminus \mathcal{P}_1'\right)\right| - \left|\mathcal{B}(\mathcal{N})\right|\right) \\
&= \lim_{n \to \infty} \frac{1}{n}\left(\left|S_{in}^{PD,\text{anti-degr}}\right|\right).
\end{aligned} \tag{S.67}$$

As follows, the available set over an anti-degradable PD channel is

$$\begin{aligned}
\left|S_{in}^{PD,\text{anti-degr}}\right| &= \left|S_{in}^{\text{degr}}\right| + \left|\mathcal{P}_1(\mathcal{N})\right| - \left|\mathcal{B}(\mathcal{N})\right| \\
&= \left|\mathcal{G}\left(\mathcal{N}_{amp},\beta\right) \cap \mathcal{G}\left(\mathcal{N}_{phase},\beta\right)\right| + \left|\mathcal{P}_1'(\mathcal{N})\right| - \left|\mathcal{B}\left(\mathcal{N}_{amp},\beta\right) \cap \mathcal{B}\left(\mathcal{N}_{phase},\beta\right)\right| \\
&= \left|S_{in}^{\text{anti-degr}}\right| + \Delta,
\end{aligned} \tag{S.68}$$

which clearly demonstrates the gain of an anti-degradable PD channel over a non-PD anti-degradable channel for boosting up the rate of quantum communication.

The code construction process resulted in five disjoint sets, $S_{in}^{PD,\text{anti-degr}}(\mathcal{N})$, $\mathcal{P}_1(\mathcal{N}) \setminus \mathcal{P}_1'(\mathcal{N})$, $\mathcal{P}_2(\mathcal{N}) \setminus \mathcal{P}_2'(\mathcal{N})$, $\mathcal{P}_2'(\mathcal{N})$ and $\mathcal{B}(\mathcal{N})$, with relations

$$S_{in}^{PD,\text{anti-degr}}(\mathcal{N}) \cap \left(\mathcal{P}_1(\mathcal{N}) \setminus \mathcal{P}_1'(\mathcal{N})\right) \cap \left(\mathcal{P}_2(\mathcal{N}) \setminus \mathcal{P}_2'(\mathcal{N})\right) \cap \mathcal{B}(\mathcal{N}) = \varnothing, \tag{S.69}$$

$$S_{in}^{PD,\text{anti-degr}}(\mathcal{N}) \cup \left(\mathcal{P}_1(\mathcal{N}) \setminus \mathcal{P}_1'(\mathcal{N})\right) \cup 2\mathcal{B}(\mathcal{N}) = [n]. \tag{S.70}$$

For the sets of valuable information and frozen bits, the relations

$$S_{in}^{PD,\text{anti-degr}}(\mathcal{N}) \cap \left(\mathcal{P}_1(\mathcal{N}) \setminus \mathcal{P}_1'(\mathcal{N})\right) \cap \left(\mathcal{P}_2(\mathcal{N}) \setminus \mathcal{P}_2'(\mathcal{N})\right) \cap \mathcal{P}_2'(\mathcal{N}) \cap \mathcal{B} = \varnothing \tag{S.71}$$

and

$$\begin{aligned}
&S_{in}^{PD,degr}\left(\mathcal{N}\right)\cup\left(\mathcal{P}_1\left(\mathcal{N}\right)\setminus\mathcal{P}_1'\left(\mathcal{N}\right)\right)\cup\left(\mathcal{P}_2\left(\mathcal{N}\right)\setminus\mathcal{P}_2'\left(\mathcal{N}\right)\right)\cup\mathcal{P}_2'\left(\mathcal{N}\right)\cup\mathcal{B}\left(\mathcal{N}\right)\\
&=\left|S_{in}^{PD,anti\text{-}degr}\right|+\left|\mathcal{P}_1\left(\mathcal{N}\right)\setminus\mathcal{P}_1'\left(\mathcal{N}\right)\right|+\left|\mathcal{P}_2\left(\mathcal{N}\right)\setminus\mathcal{P}_2'\left(\mathcal{N}\right)\right|+\left|\mathcal{P}_2'\left(\mathcal{N}\right)\right|+2\mathcal{B}\left(\mathcal{N}\right)\\
&=\left|S_{in}^{PD,anti\text{-}degr}\right|+\left|\mathcal{P}_1\left(\mathcal{N}\right)\setminus\mathcal{P}_1'\left(\mathcal{N}\right)\right|+\left|\mathcal{P}_2\left(\mathcal{N}\right)\right|+2\mathcal{B}\left(\mathcal{N}\right)\\
&=\left|S_{in}^{PD,anti\text{-}degr}\right|+\left|\mathcal{P}_1\left(\mathcal{N}\right)\setminus\mathcal{P}_1'\left(\mathcal{N}\right)\right|+2\mathcal{B}\left(\mathcal{N}\right)\\
&=[n]=\left[2^k\right]
\end{aligned} \quad (S.72)$$

holds for an anti-degradable PD channel.

*These results conclude the proof of Theorem S.2.*

∎

## S.1.4 Rate of entanglement consumption

**Corollary S.1** (On the net rate of a degradable PD channel). *For a degradable PD channel $\mathcal{N}$, the rate of entanglement consumption is zero. The net rate is equal to the rate of quantum communication $R_Q(\mathcal{N})$. The rate $R_Q(\mathcal{N})$ will achieve $Q(\mathcal{N})$ by the use of set $S_{in}^{PD,degr}(\mathcal{N})$.*

**Corollary S.2** (On the net rate of an anti-degradable PD channel). *For an anti-degradable PD channel the rate of entanglement consumption is non-zero, and equal to the standard anti-degradable channels. The net rate will be equal to the rate of quantum communication $R_Q(\mathcal{N})$ minus the rate of entanglement consumption. The rate $R_Q(\mathcal{N})$ will achieve $Q(\mathcal{N})$ by the use of set $S_{in}^{PD,anti\text{-}degr}(\mathcal{N})$.*

**Theorem S.3** *The quantum capacity of a PD channel cannot be achieved by those polar codes that are defined for non-PD channels.*

*Proof.*
The proof trivially follows from the previously obtained results for the degradable PD and for the anti-degradable PD channel. Since $\lim_{n\to\infty}\frac{1}{n}\left(\left|S_{in}^{degr}(\mathcal{N})\right|\right)<\lim_{n\to\infty}\frac{1}{n}\left(\left|S_{in}^{PD,degr}(\mathcal{N})\right|\right)$, where $\lim_{n\to\infty}\frac{1}{n}\left(\left|S_{in}^{PD,degr}(\mathcal{N})\right|\right)=\max_{\forall\rho_i}R_Q(\mathcal{N})\approx\max_{\forall\rho_i}\left(I_{coh}(\mathcal{N})\right)$, it follows that for a degradable PD channel $\mathcal{N}$, $\lim_{n\to\infty}\frac{1}{n}\left(\left|S_{in}^{degr}(\mathcal{N})\right|\right)\ll Q(\mathcal{N})$. For an anti-degradable PD channel $\mathcal{N}$, we have

$$\lim_{n\to\infty} \frac{1}{n}\left(\left|S_{in}^{non-degr}\left(\mathcal{N}\right)\right|\right) \ll Q(\mathcal{N}),$$ where $S_{in}^{degr}$ and $S_{in}^{anti\text{-}degr}$ are the sets constructed for degradable non-PD and anti-degradable non-PD channels.

*These results conclude the proof of Theorem S.3.*

∎

## S.1.5 Summary on Code Construction

In this section we summarize the main results on the polar coding scheme used for a PD channel.

### A degradable channel

If channel $\mathcal{N}$ is degradable but not PD, then only the set $S_{in}^{degr}\left(\mathcal{N}\right) = \mathcal{G}\left(\mathcal{N}_{ampl},\beta\right) \cap \mathcal{G}\left(\mathcal{N}_{phs},\beta\right)^{degr}$ can be used for quantum communication. This will result in the quantum data rate

$$\begin{aligned} R_Q\left(\mathcal{N}\right) &= \lim_{n\to\infty} \frac{1}{n}\left(\left|S_{in}^{degr}\left(\mathcal{N}\right)\right|\right) \\ &= \lim_{n\to\infty} \frac{1}{n}\left|\left(\mathcal{G}\left(\mathcal{N}_{amp},\beta\right) \cap \mathcal{G}\left(\mathcal{N}_{phase},\beta\right)\right)^{degr}\right|. \end{aligned} \tag{S.73}$$

The codeword sets with the achievable quantum rates for this channel $\mathcal{N}$ are depicted in grey in Fig. S.2.

|  | $\mathcal{P}_1$ | $S_{in}^{degr}$ |  |
|---|---|---|---|
|  | $R_Q\left(\mathcal{N}_{AB}\right) = 0$ | $R_Q\left(\mathcal{N}_{AB}\right) > 0$ | $\mathcal{G}\left(\mathcal{N}_{ampl},\beta\right)$ |
|  | $\mathcal{B}$ | $\mathcal{P}_2$ |  |
|  | $R_Q\left(\mathcal{N}_{AB}\right) = 0$ | $R_Q\left(\mathcal{N}_{AB}\right) = 0$ | $\mathcal{B}\left(\mathcal{N}_{ampl},\beta\right)$ |
|  | $\mathcal{B}\left(\mathcal{N}_{phs},\beta\right)$ | $\mathcal{G}\left(\mathcal{N}_{phs},\beta\right)$ |  |

**Figure S.2.** The information of logical channel $\mathcal{N}_{AB}$ for the degradable but not PD channel $\mathcal{N}$. The codewords that can transmit amplitude information are depicted by $\mathcal{G}\left(\mathcal{N}_{amp},\beta\right)$. The set that can transmit phase is denoted by $\mathcal{G}\left(\mathcal{N}_{phase},\beta\right)$. For a degradable $\mathcal{N}$, only the set $\left|S_{in}^{degr}\left(\mathcal{N}\right)\right| = k < l$ can be used for quantum communication.

For a degradable channel $\mathcal{N}$, the valuable amplitude information can be leaked only from the set $\mathcal{P}_1(\mathcal{N})$, which finally results in $\left(S_{in}^{degr}(\mathcal{N}) \cup \mathcal{P}_1(\mathcal{N}) \cup \mathcal{P}_2(\mathcal{N})\right) \setminus \left(\mathcal{P}_1(\mathcal{N}) \cup \mathcal{P}_2(\mathcal{N})\right)$, where $\lim_{n \to \infty} \frac{1}{n}|\mathcal{P}_2(\mathcal{N})| = 0$, i.e., $\left(S_{in}^{degr}(\mathcal{N}) \cup \mathcal{P}_1(\mathcal{N})\right) \setminus \left(\mathcal{P}_1(\mathcal{N})\right)$. The codeword set $\mathcal{P}_1(\mathcal{N})$ represents the information from complementary channel $\mathcal{N}_{AE}$ of $\mathcal{N}$, between Alice and the environment. For these sets, $R_Q(\mathcal{N}_{AE}) > 0$ and $R_Q(\mathcal{N}_{AB}) = 0$. These results are summarized in Fig. S.3.

|  | $\mathcal{P}_1$ | $S_{in}^{degr}$ |  |
|---|---|---|---|
|  | $R_Q(\mathcal{N}_{AE}) > 0$ | $R_Q(\mathcal{N}_{AE}) = 0$ | $\mathcal{G}(\mathcal{N}_{ampl}, \beta)$ |
|  | $\mathcal{B}$ | $\mathcal{P}_2$ |  |
|  | $R_Q(\mathcal{N}_{AE}) = 0$ | $R_Q(\mathcal{N}_{AE}) = 0$ | $\mathcal{B}(\mathcal{N}_{ampl}, \beta)$ |
|  | $\mathcal{B}(\mathcal{N}_{phs}, \beta)$ | $\mathcal{G}(\mathcal{N}_{phs}, \beta)$ |  |

**Figure S.3.** The information of complementary channel $\mathcal{N}_{AE}$ for the degradable channel $\mathcal{N}$. Set $\mathcal{P}_1(\mathcal{N})$ represents the information of channel $\mathcal{N}_{AE}$. For this set, $R_Q(\mathcal{N}_{AE}) > 0$ and $R_Q(\mathcal{N}_{AB}) = 0$. Set $\mathcal{P}_2(\mathcal{N})$ is empty.

If $\mathcal{N}$ is assumed to be only degradable, then only the set $S_{in}^{degr}(\mathcal{N})$ can be used for quantum communication, due to the properties of the complementary channels $\mathcal{N}_{AE}$ and $\mathcal{N}_{AE'}$ of $\mathcal{N}$.

**A degradable PD channel**

By exploiting these properties of channel $\mathcal{N}$, better quantum communication rates can be achieved, in comparison to the case where $\mathcal{N}$ is assumed to be only degradable.

The achievable codeword set for quantum communication has been increased to $S_{in}^{PD,degr}(\mathcal{N}) = S_{in}^{degr}(\mathcal{N}) \cup \mathcal{P}_1'(\mathcal{N}) \cup \mathcal{P}_2'(\mathcal{N})$, which results in the quantum data transmission rate

$$\begin{aligned} R_Q(\mathcal{N}) &= \lim_{n\to\infty}\frac{1}{n}\left(\left|S_{in}^{degr}(\mathcal{N})\cup \mathcal{P}_1'(\mathcal{N})\cup \mathcal{P}_2'(\mathcal{N})\right|\right) \\ &= \lim_{n\to\infty}\frac{1}{n}\left|\left(\mathcal{G}(\mathcal{N}_{ampl},\beta)\cap \mathcal{G}(\mathcal{N}_{phs},\beta)\right)^{degr}\cup \mathcal{P}_1'(\mathcal{N})\cup \mathcal{P}_2'(\mathcal{N})\right| \\ &= \lim_{n\to\infty}\frac{1}{n}\left(\left|S_{in}^{degr}(\mathcal{N})\cup \mathcal{P}_1'(\mathcal{N})\right|\right) \hspace{2cm} (S.74)\\ &= \lim_{n\to\infty}\frac{1}{n}\left|\left(\mathcal{G}(\mathcal{N}_{ampl},\beta)\cap \mathcal{G}(\mathcal{N}_{phs},\beta)\right)^{degr}\cup \mathcal{P}_1'(\mathcal{N})\right| \\ &= \lim_{n\to\infty}\frac{1}{n}\left(\left|S_{in}^{PD,degr}(\mathcal{N})\right|\right), \end{aligned}$$

where $\lim_{n\to\infty}\frac{1}{n}|\mathcal{P}_2(\mathcal{N})|=0$, $\lim_{n\to\infty}\frac{1}{n}|\mathcal{P}_2(\mathcal{N})\setminus \mathcal{P}_2'(\mathcal{N})|=0$ and $\lim_{n\to\infty}\frac{1}{n}|\mathcal{P}_2'(\mathcal{N})|=0$. Our results on a degradable PD channel $\mathcal{N}$ are summarized in Fig. S.4.

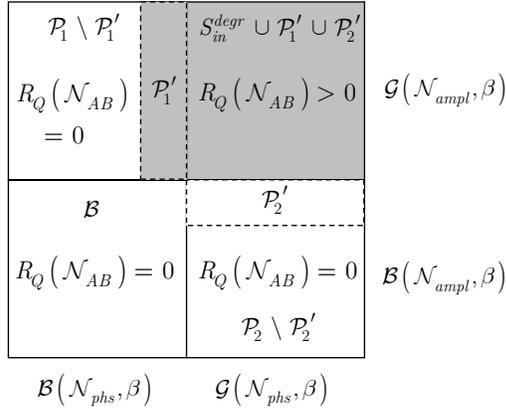

**Figure S.4**. The information of logical channel $\mathcal{N}_{AB}$ for a degradable PD channel $\mathcal{N}$. For the PD channel the extended set $S_{in}^{PD,degr}(\mathcal{N}) = S_{in}^{degr}(\mathcal{N})\cup \mathcal{P}_1'(\mathcal{N})\cup \mathcal{P}_2'(\mathcal{N})$ with $\left|S_{in}^{PD,degr}(\mathcal{N})\right| = \left|S_{in}^{degr}(\mathcal{N})\right| + \Delta = k + \Delta$ will be available, where $\mathcal{P}_1'(\mathcal{N})\subseteq \mathcal{P}_1(\mathcal{N})$, $\mathcal{P}_2'(\mathcal{N})\subseteq \mathcal{P}_2(\mathcal{N})$, $\lim_{n\to\infty}\frac{1}{n}|\mathcal{P}_2(\mathcal{N})\setminus \mathcal{P}_2'(\mathcal{N})| = 0$, and $\lim_{n\to\infty}\frac{1}{n}|\mathcal{P}_2'(\mathcal{N})| = 0$. Bob synthesizes set $\mathcal{P}_2$ by the corresponding subset $\Omega_{\mathcal{P}_1\setminus \mathcal{P}_1'}$ of $(\mathcal{P}_1\setminus \mathcal{P}_1')$.

The extended set will result in higher quantum communication rates, in comparison to the non-PD, degradable-only channel $\mathcal{N}$. Since the defined channel $\mathcal{N}$ is degradable and also PD, the amount of information in the degraded environment state $E'$ outputted by channel $\mathcal{N}_{AE'} = \mathcal{N}_{AE}\circ \mathcal{D}^{E\to E'}$ is less than that outputted by channel $\mathcal{N}_{AE}$ in the environment state $E$. Valuable information can be leaked to the environment only from the set $\mathcal{P}_1(\mathcal{N})\setminus \mathcal{P}_1'(\mathcal{N})$, where $\mathcal{P}_1'(\mathcal{N})\subseteq \mathcal{P}_1(\mathcal{N})$. These sets are depicted in Fig. S.5.

| $\mathcal{P}_1 \setminus \mathcal{P}_1'$ | $S_{in}^{degr} \cup \mathcal{P}_1' \cup \mathcal{P}_2'$ | |
|---|---|---|
| $R_Q(\mathcal{N}_{AE'})$ $> 0$   $\mathcal{P}_1'$ | $R_Q(\mathcal{N}_{AE'}) = 0$ | $\mathcal{G}(\mathcal{N}_{ampl}, \beta)$ |
| $\mathcal{B}$ | $\mathcal{P}_2'$ | |
| $R_Q(\mathcal{N}_{AE'}) = 0$ | $R_Q(\mathcal{N}_{AE'}) = 0$ $\mathcal{P}_2 \setminus \mathcal{P}_2'$ | $\mathcal{B}(\mathcal{N}_{ampl}, \beta)$ |
| $\mathcal{B}(\mathcal{N}_{phs}, \beta)$ | $\mathcal{G}(\mathcal{N}_{phs}, \beta)$ | |

**Figure S.5**. The information of the degraded complementary channel $\mathcal{N}_{AE'} = \mathcal{N}_{AE} \circ \mathcal{D}^{E \to E'}$ for the degradable PD channel $\mathcal{N}$. For the degradable PD channel $\mathcal{N}$, set $\mathcal{P}_1'(\mathcal{N}) \setminus \mathcal{P}_1(\mathcal{N})$ represents the information of channel $\mathcal{N}_{AE'}$. For this set, $R_Q(\mathcal{N}_{AE'}) > 0$ and $R_Q(\mathcal{N}_{AB}) = 0$. The sets $\mathcal{P}_2'(\mathcal{N})$ and $\mathcal{P}_2'(\mathcal{N}) \setminus \mathcal{P}_2(\mathcal{N})$ are empty.

**An anti-degradable channel**

For an anti-degradable channel the set $S_{in}^{anti\text{-}degr}(\mathcal{N}) = \left(\mathcal{G}(\mathcal{N}_{amp}, \beta) \cap \mathcal{G}(\mathcal{N}_{phase}, \beta)\right)^{degr} \setminus \mathcal{B}$ can be used for quantum communication. This will result in the quantum data rate

$$R_Q(\mathcal{N}) = \lim_{n \to \infty} \frac{1}{n} \left( \left| S_{in}^{degr}(\mathcal{N}) \right| - \left| \mathcal{B}(\mathcal{N}) \right| \right) \\ = \lim_{n \to \infty} \frac{1}{n} \left| S_{in}^{anti\text{-}degr}(\mathcal{N}) \right|. \tag{S.75}$$

**An anti-degradable PD channel**

The achievable codeword set for quantum communication has been increased from $S_{in}^{anti\text{-}degr}(\mathcal{N})$ to $S_{in}^{PD,anti\text{-}degr}(\mathcal{N}) = S_{in}^{anti\text{-}degr}(\mathcal{N}) \cup \mathcal{P}_1'(\mathcal{N}) \cup \mathcal{P}_2'(\mathcal{N})$, which results in the quantum data transmission rate

$$\begin{aligned}
R_Q(\mathcal{N}) &= \lim_{n\to\infty} \frac{1}{n}\left(\left|S_{in}^{degr}(\mathcal{N}) \cup \mathcal{P}_1'(\mathcal{N}) \cup \mathcal{P}_2'(\mathcal{N})\right| - |\mathcal{B}(\mathcal{N})|\right) \\
&= \lim_{n\to\infty} \frac{1}{n}\left(\left|\left(\mathcal{G}(\mathcal{N}_{ampl},\beta) \cap \mathcal{G}(\mathcal{N}_{phs},\beta)\right)^{degr} \cup \mathcal{P}_1'(\mathcal{N}) \cup \mathcal{P}_2'(\mathcal{N})\right| - |\mathcal{B}(\mathcal{N})|\right) \\
&= \lim_{n\to\infty} \frac{1}{n}\left(\left|S_{in}^{anti\text{-}degr}(\mathcal{N}) \cup \mathcal{P}_1'(\mathcal{N})\right|\right) \\
&= \lim_{n\to\infty} \frac{1}{n}\left|\left(\mathcal{G}(\mathcal{N}_{ampl},\beta) \cap \mathcal{G}(\mathcal{N}_{phs},\beta)\right)^{degr} \setminus \mathcal{B}(\mathcal{N}) \cup \mathcal{P}_1'(\mathcal{N})\right| \qquad (\text{S.76}) \\
&= \lim_{n\to\infty} \frac{1}{n}\left(\left|S_{in}^{anti\text{-}degr}(\mathcal{N})\right| + |\mathcal{P}_1'(\mathcal{N})|\right) \\
&= \lim_{n\to\infty} \frac{1}{n}\left(\left|S_{in}^{PD,degr}(\mathcal{N})\right| - |\mathcal{B}(\mathcal{N})|\right) \\
&= \lim_{n\to\infty} \frac{1}{n}\left(\left|S_{in}^{PD,anti\text{-}degr}(\mathcal{N})\right|\right),
\end{aligned}$$

since, for an anti-degradable PD channel $\lim_{n\to\infty}\frac{1}{n}|\mathcal{P}_2(\mathcal{N})| = 0$, $\lim_{n\to\infty}\frac{1}{n}\left|\mathcal{P}_2(\mathcal{N}) \setminus \mathcal{P}_2'(\mathcal{N})\right| = 0$ and $\lim_{n\to\infty}\frac{1}{n}|\mathcal{P}_2'(\mathcal{N})| = 0$.